\documentclass[a4paper,aps,prl,showpacs, twocolumn, superscriptaddress]{revtex4-2} 

\usepackage[utf8]{inputenc}  
\usepackage[T1]{fontenc}     
\usepackage[american]{babel}  
\usepackage[colorlinks,citecolor=blue,linkcolor=magenta,urlcolor=blue]{hyperref}  
\usepackage{graphicx} 
\usepackage{amsmath,amssymb,amsthm,bm,mathtools,amsfonts,mathrsfs,bbm,dsfont} 

\newcommand{\bra}[1]{\langle #1|}
\newcommand{\ket}[1]{|#1\rangle}

\newcommand{\ketbra}[1]{| #1\rangle \langle #1|}

\newcommand{\be}{\begin{equation}}
\newcommand{\ee}{\end{equation}}
\newcommand{\bea}{\begin{eqnarray}}
\newcommand{\eea}{\end{eqnarray}}

\newcommand{\mean}[1]{\ensuremath{\langle{#1}\rangle}}

\newcommand{\eins}{\mathbbm{1}}

\newcommand{\RR}{\ensuremath{\mathcal{R}}}

\newcommand{\SSS}{\ensuremath{\mathcal{S}}}

\newcommand{\kommentar}[1]{}
\newcommand{\tr}{{\mathrm{tr}}}

\renewcommand{\vr}{\ensuremath{\varrho}}

\newcommand{\forget}[1]{}

\usepackage{graphicx, xcolor}
\usepackage{braket}

\theoremstyle{remark}
\newtheorem*{remark}{Remark}
\begin{document}

\title{ Bound entanglement from randomized measurements
}
\author{Satoya Imai}
\affiliation{Naturwissenschaftlich-Technische Fakult\"at, Universit\"at Siegen, Walter-Flex-Str.~3, D-57068 Siegen, Germany}

\author{Nikolai Wyderka}
\affiliation{Institut für Theoretische Physik III, Heinrich-Heine-Universität Düsseldorf, Universitätsstr.~1, D-40225 Düsseldorf, Germany}
\affiliation{Naturwissenschaftlich-Technische Fakult\"at, Universit\"at Siegen, Walter-Flex-Str.~3, D-57068 Siegen, Germany}

\author{Andreas Ketterer}
\affiliation{Physikalisches Institut, Albert-Ludwigs-Universit\"at Freiburg, Hermann-Herder-Str. 3, D-79104 Freiburg, Germany}
\affiliation{EUCOR Centre for Quantum Science and Quantum Computing, Albert-Ludwigs-Universit\"at Freiburg, Hermann-Herder-Str.~3, D-79104~Freiburg, Germany}

\author{Otfried Gühne}
\affiliation{Naturwissenschaftlich-Technische Fakult\"at, Universit\"at Siegen, Walter-Flex-Str.~3, D-57068 Siegen, Germany}

\date{\today}

\begin{abstract}
If only limited control over a multiparticle quantum system is available, 
a viable method to characterize correlations is to perform random measurements 
and consider the moments of the resulting probability distribution. We present
systematic methods to analyze the different forms of entanglement with these
moments in an optimized manner. First, we find the optimal criteria for different 
forms of multiparticle entanglement in three-qubit systems using the second moments
of randomized measurements. Second, we present the optimal inequalities if 
entanglement in a bipartition of a multi-qubit system shall be analyzed in terms 
of these moments. Finally, for higher-dimensional two-particle systems and higher 
moments, we provide criteria that are able to characterize various examples of 
bound entangled states, showing that detection of such states is possible in this framework.
\end{abstract}

\maketitle

{\it Introduction.---}
With the current development of experimental quantum technologies, larger
quantum systems with more and more particles become available, but 
controlling and analyzing these systems is complicated. In fact, due
to the exponentially increasing dimension of the underlying Hilbert 
space, a complete characterization of the quantum states or quantum 
dynamics is quickly out of reach. A key idea for analyzing large
quantum systems is therefore to perform {\it random} measurements or
operations, and to characterize the global quantum system with the help
of the observed statistics. Examples are procedures like randomized
benchmarking for the analysis of quantum gates \cite{Emerson2005, Knill2008}, 
certain methods for estimating the fidelity of quantum states 
\cite{Flammia2011}, or various proposals to perform variants of state tomography 
using random measurements \cite{Aaronson2007, Huang2020, Morris2019, Brandao2020}.

It was noted early that randomized measurements could also be used
to study quantum correlations \cite{Liang2010, Wallman2012, Shadbolt2012}.
The original motivation came from the situation where two parties, typically 
called Alice and Bob, share a quantum state, but no common reference frame.
This situation has been discussed in a variety of settings in quantum information
processing \cite{Bartlett2003, Bartlett2007, Laing2010, Hayden2017}.
Although the determination of the entire quantum state is impossible in this setting,
it may still be analyzed along the following lines.
Alice and Bob perform separate measurements, denoted by $M_A$ and $ M_B$, and rotate them arbitrarily. That is, they evaluate an expression of the form
\be
\mean{M_{A}\otimes M_B}_{U_{A}\otimes U_B} = \mathrm{tr}[\varrho_{AB} (U_A M_A U_A^\dagger)\otimes (U_B M_B U_B^\dagger)],
\label{eq-onemeasurement}
\ee
which, of course, depends on the chosen unitary ${U_{A}\otimes U_B}$. The prime
idea is to sample random unitaries and consider the resulting 
probability distribution of $\mean{M_{A}\otimes M_B}_{U_{A}\otimes U_B}$. 
This probability distribution contains valuable information about the state, 
and the distribution may be characterized by its moments
\be
\RR^{(r)}_{AB} = \int dU_A \int dU_B [\mean{M_{A}\otimes M_B}_{U_{A}\otimes U_B}]^r,
\label{eq-moments1}
\ee
where the unitaries are typically chosen according to the Haar 
distribution. Clearly, similar moments can be defined for multiparticle 
systems.

In recent years, several works proceeded in this direction. One research 
line has been started from the estimation of the state's purity 
\cite{vanEnk2012}, and then protocols for measuring entanglement via 
R\'{e}nyi entropies have been presented \cite{Elben2018} and experimentally 
implemented \cite{Brydges2019}. Very recently, ideas to estimate the 
entanglement criterion of the positivity of the partial transpose (PPT) 
\cite{Peres1996, PHorodecki1996} have been introduced \cite{Zhou2020, Elben2020}. 
Another research line characterized the relation of the second 
moments \cite{Tran2015,Tran2016} $\RR^{(2)}_{AB}$
and those of the marginals \cite{Knips2020} to entanglement. Recently, 
higher moments have been used to characterize multiparticle entanglement
\cite{Ketterer2019, Ketterer2020a}, and quantum designs have been shown to 
allow for a simplified implementation, as the integral in Eq.~(\ref{eq-moments1}) 
can be replaced by finite sums \cite{Ketterer2019, KettererarXiv, Knips2020a}.

Still, the present results along the above research lines are incomplete in several respects.
First, while many entanglement criteria have been presented, their optimality is not
clear. It would be desirable to use the information obtained by 
randomized measurements most efficiently. Second, the known results from randomized measurements allow one to detect highly entangled states only, e.g., states
which are close to
 pure states. For a long-range impact of the research program, 
however, it is vital that also weakly entangled states, e.g., the ones that cannot be detected by the PPT criterion, can be analyzed.

The goal of this paper is to generalize the existing approaches in
two directions: First, we will systematically consider the moments 
of the measurement results when only some of the parties measure. That is, we 
evaluate the expressions in Eqs.~(\ref{eq-onemeasurement},
\ref{eq-moments1}) for the special case of $M_A=\eins$ or $M_B=\eins$,
and call these quantities the reduced moments $\RR^{(r)}_B$
and $\RR^{(r)}_A$. Note that this case effectively corresponds to discarding
the measurements of Alice for $\RR^{(r)}_B$ (resp., Bob for
$\RR^{(r)}_A$), such that the reduced moments can directly be evaluated 
from the data taken for measuring $\RR^{(r)}_{AB}$. As we will
show, in terms of these reduced moments, improved entanglement criteria
can be designed, which are optimal in the sense that if a quantum state 
is not detected by them, then there is also a separable state compatible 
with the data.

Second, we present a systematic approach to characterize high-dimensional
systems with higher moments $\RR^{(r)}_{AB}$. We show how previously known 
entanglement criteria \cite{Rudolph2005, Chen2003, deVicente2007} can be 
formulated in terms of moments. With this, we demonstrate that bound 
entanglement, a weak form of entanglement that cannot be used for 
entanglement distillation and is not detectable by the PPT criterion, can 
be also characterized in a reference-frame independent manner. This
shows
that the approach of randomized measurements is powerful 
enough to characterize the rich plethora of entanglement phenomena.

{\it Multiparticle correlations for three qubits.---}
For three qubits,  the measurements
$M_A, M_B$, and $M_C$ may, without loss of generality, be taken as the Pauli-$Z$
matrix $\sigma_3$.
If we consider the full or reduced second moments $\RR^{(2)}$, 
the analysis is simplified by the fact that each integral over 
$U(2)$ in Eq.~(\ref{eq-moments1}) can be replaced by sums over
the four Pauli matrices $\sigma_0, \sigma_1, \sigma_2,$ and $\sigma_3$.
This is because the Pauli matrices form a unitary two-design,
meaning that averages of polynomials of degree two or less yield
identical results \cite{Dankert2005, Ketterer2019}.

Accordingly, the moments are directly related to the Bloch decomposition of
the three-qubit state $\varrho_{ABC}$. Recall that any three-qubit state can
be written as
\be 
\label{eq-threequbit}
\varrho_{ABC} = \frac{1}{8} \sum_{i,j,k =0}^{3} \alpha_{ijk} 
\sigma_i \otimes \sigma_j \otimes\sigma_k,
\ee
where $\sigma_0 = \eins$ denotes the identity matrix. The full and reduced
second moments can simply be expressed in terms of the coefficients
$\alpha_{ijk}$, and they read
\begin{align} 
\RR^{(2)}_{ABC}  = 
\frac{1}{27}\!\!\!\!\sum_{i,j,k=1}^3 \!\!\!\!\alpha_{ijk}^2,
\;
\RR^{(2)}_{AB}  = \frac{1}{9}\!\!\sum_{i,j=1}^3 \!\!\alpha_{ij0}^2,
\;
\RR^{(2)}_{A} = \frac{1}{3}\!\!\sum_{i=1}^3 \!\!\alpha_{i00}^2,
\label{eq-secmoments-three}
\end{align}
and similarly for the reduced moments on other parts of the three-particle
system. 

\begin{figure}[t]
    \centering
    \includegraphics[width=0.9\columnwidth]{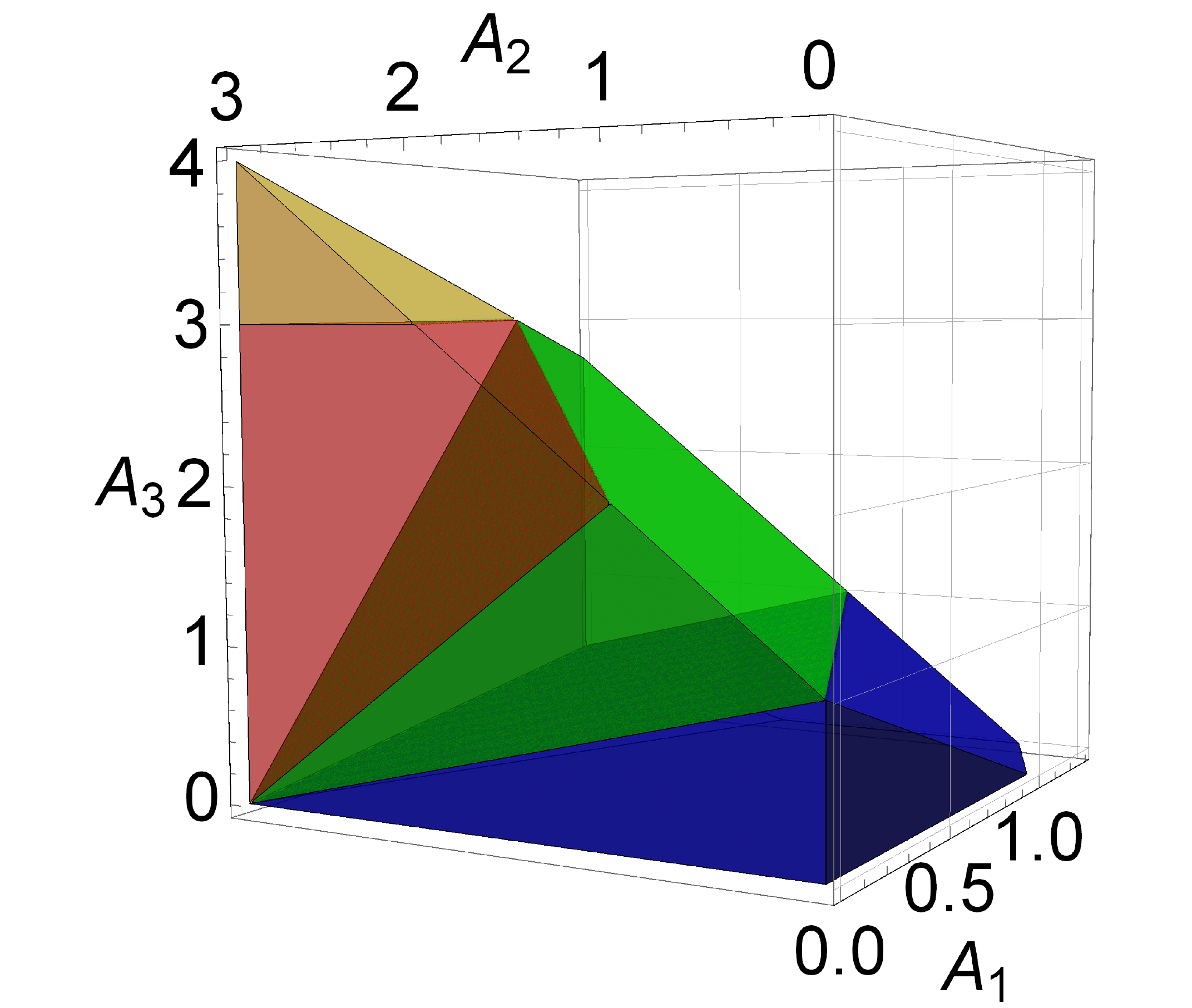}
    \caption{
Geometry of the three-qubit state space in terms of the second moments of
random measurements or sector lengths. The total polytope is the set of all 
states, characterized by the inequalities $A_k \geq 0$, $A_1-A_2+A_3 \leq 1$,
$A_2 \leq 3$, and $A_1 + A_2 \leq 3(1+A_3)$ \cite{Nikolai2020}. The fully 
separable states are contained in the blue polytope, obeying the additional
constraint in Eq.~(\ref{obs1-fs}) in Observation 1. States that are biseparable for some 
partitions are contained in the union of the green and blue polytopes, 
characterized by the additional constraint in Eq.~(\ref{obs2-bs}) from Observation 2. In 
fact, for any point in the green and blue areas, there is a biseparable
state with the corresponding second moments. The yellow area corresponds to 
the states violating the best previously known criterion for biseparable 
states, $A_3 \leq 3$ \cite{Aschauer2003, Vicente2011, Klockl2015, Nikolai2020}.
Thus, the red area marks the improvement of the criterion in Observation 2 compared 
with previous results.
}
    \label{threequbits}
\end{figure}

Sums of this form have already been considered under the concept of
sector lengths \cite{Aschauer2003, Vicente2011, Klockl2015, Eltschka2020, 
Nikolai2020} and multiparticle concurrences \cite{Mintert2005,Aolita2008}.
More precisely, the notion of sector lengths captures the magnitude
of the one-, \mbox{two-,} and three-body correlations in the state
$\varrho_{ABC}$, where the
one- and two-body correlations are averaged over all one- and two-particle
reduced states. That is, the sector lengths $A_k$ are given by
$A_1 = 3(\RR^{(2)}_{A}+\RR^{(2)}_{B}+\RR^{(2)}_{C})$,
$A_2 = 9(\RR^{(2)}_{AB}+\RR^{(2)}_{AC}+\RR^{(2)}_{BC})$,
and 
$A_3 = 27\RR^{(2)}_{ABC}$. Most importantly, the set of all three-qubit
states forms a polytope in the space of the sector lengths,  which has
recently been fully characterized \cite{Nikolai2020}, see also
Fig.~\ref{threequbits}.

To proceed, recall that a state is fully separable if it can be written 
as
\be 
\label{eq-fs}
\varrho_{\text{fs}} = \sum_k p_k \vr^A_k \otimes \vr^B_k \otimes \vr^C_k,
\ee
where the $p_k$ form a probability distribution. Now, we can formulate the first main 
result of this paper.

{\bf Observation 1.} 
{\it Any fully separable three-qubit state obeys
\begin{align} 
\label{obs1-fs}
A_2 + 3A_3 \leq 3 +A_1
\end{align}
or, equivalently,
$ 
3(\RR^{(2)}_{AB}+\RR^{(2)}_{AC}+\RR^{(2)}_{BC})
+
27\RR^{(2)}_{ABC}
\leq 
1 + \RR^{(2)}_{A}+\RR^{(2)}_{B}+\RR^{(2)}_{C}
$.
This is the optimal linear criterion in the sense that any other linear 
criterion for the $A_i$ detects strictly fewer states.}

The proof of this Observation, including possible generalizations
to higher-dimensional systems, is given in Appendix A in the Supplemental
Material (SM) \cite{SM}, and the
geometrical interpretation is displayed in Fig.~\ref{threequbits}.

Violation of  Eq.~(\ref{obs1-fs}) implies that the 
state contains some entanglement, but it does not mean that all 
three particles are entangled. Indeed, an entangled state may still 
be separable with respect to some bipartition.  For instance, if we 
consider the bipartition $A|BC$, a state separable with respect to 
this bipartition can be written as
$
\varrho_{{A|BC}} = \sum_k q^A_k \vr^A_k \otimes \vr^{BC}_k,
$
where the $q^A_k$ form a probability distribution and $\vr^{BC}_k$
may be entangled. Similarly, one can define biseparable states with
respect to the two other bipartitions as $\varrho_{{B|AC}}$ and
$\varrho_{{C|AB}}$. For these states, we can formulate:

{\bf Observation 2.} 
{\it Any three-qubit state which is
separable with respect to some bipartition obeys
\begin{align} \label{obs2-bs}
A_2 + A_3 \leq 3 (1+ A_1)
\end{align}
or, equivalently,
$ 
3(\RR^{(2)}_{AB}+\RR^{(2)}_{AC}+\RR^{(2)}_{BC})
+
9\RR^{(2)}_{ABC}
\leq 
1 + 3(\RR^{(2)}_{A}+\RR^{(2)}_{B}+\RR^{(2)}_{C})
$.
This is the optimal criterion in the sense that
if the three $A_i$ obey the inequality, then for 
any bipartition there is a separable state 
compatible  with them. 
}

Again, the proof and the generalizations to higher dimensions are
given in Appendix A \cite{SM}, and the geometry is displayed in Fig.\,\ref{threequbits}.
We add that we have strong numerical evidence that Eq.~(\ref{obs2-bs})
also holds for mixtures of biseparable states with respect to different
partitions, i.e., states of the form
$
\varrho_{\text{bs}} = p_A \varrho_{A|BC} + p_B \varrho_{B|AC} + p_C \varrho_{C|AB},
$
where the $p_A$, $p_B$, and $p_C$ form convex weights. Nevertheless, we leave this
as a conjecture for further study. More detailed information on the numerical methods used can be found in Appendix~D \cite{SM}.

Our two observations show that not only the three-body second moment
$\RR^{(2)}_{ABC}$, but also the one- and two-body reduced moments
such as $\RR^{(2)}_{AB}$ and $\RR^{(2)}_{A}$ can be useful for entanglement
detection. In fact, their linear combinations allow to detect entangled
states more efficiently than existing criteria
\cite{Aschauer2003, Vicente2011, Klockl2015, Nikolai2020},
see also Appendix A \cite{SM}.

In particular, as shown in the Appendix, Eq.~(\ref{obs2-bs})
can detect multipartite entanglement for mixtures of
Greenberger-Horne-Zeilinger (GHZ) states and W states
(i.e., $\ket{\text{GHZ}} = \frac{1}{\sqrt{2}} (\ket{000}+\ket{111})$,
$\ket{\text{W}}=\frac{1}{\sqrt{3}}(\ket{001}+\ket{010}+\ket{100})$, 
even if two other important entanglement measures, namely the 
three-tangle and bipartite entanglement in the reduced 
subsystems vanish simultaneously \cite{Coffman2000}.

{\it Optimal criteria for general bipartitions.---}
In many realistic scenarios, it is sufficient to detect entanglement across some fixed bipartition $I|\bar{I}$ of the multiparticle system. For this task, second moments of randomized measurements can be used as well: Performing random measurements at each qubit and 
considering the second moments allows one to generalize the moments 
in Eq.~(\ref{eq-secmoments-three}) for the given number of qubits. In turn, these moments allow one to determine the quantities $\tr(\vr_I^2)$, $\tr(\vr_{\bar{I}}^2)$, and $\tr(\vr^2)$ for the reduced states of the 
bipartition and the global state. This approach has recently been used in an experiment \cite{Brydges2019}, where entanglement criteria with  the second-order R\'{e}nyi entropy $S_2(\varrho_X) = -\log_2{\tr(\varrho_X^2)}$ were employed. The entropic criteria for separable states 
read  $S_2(\varrho_X) \leq  S_2(\varrho)$ for $X=I, \bar{I}$; if this is violated, then $\varrho$ is entangled 
\cite{HorodeckiMajori1996, Nielsen2001, Elben2018}. 

Using our methods, we can show that this approach is optimal.
To formulate the result, we assume that both sides of
the bipartition have the same number of qubits.
Then, recall 
that any bipartite state can be written as 
\be 
\label{eq-twoqudits}
\varrho_{AB} = 
\frac{1}{d^{2}} \sum_{i,j=0}^{d^2-1} t_{ij} \lambda_i \otimes \lambda_j,
\ee
where $\lambda_{0}=\eins$ denotes the identity matrix and $\lambda_{i}$ are the Gell-Mann matrices \cite{Kimura2003, Bertlmann2008}.
This is the decomposition of $\varrho_{AB}$ using the basis of Hermitian, orthogonal, and traceless matrices, i.e., 
$\lambda_i=\lambda_i^\dagger$,
$\mathrm{tr}\left[\lambda_{i} \lambda_{j}\right]=d \delta_{i j}$,
and $\mathrm{tr}\left[\lambda_{i} \right]=0$ for $i > 0$.
These properties are the natural extensions of Pauli matrices for $SU(2)$ to $SU(d)$, which are used in particle physics \cite{GellMann1962}.
The quantities of interest are
\begin{align} 
A_2& = \sum_{i,j=1}^{d^2-1} t_{ij}^2,
\;\;
A_1^{A}  = \sum_{i=1}^{d^2-1} t_{i0}^2,
\;\;
A_1^B = \sum_{i=1}^{d^2-1} t_{0i}^2.
\end{align}
We also define $A_1 = A_1^A + A_1^B$, which allows to recover the purities via $\tr(\vr_{AB}^2)=(1+A_1+A_2)/d^2$ and $\tr({\vr_A^2})=(1+A_1^A)/d.$ 
It is interesting that, although the $\lambda_i$ are {\it not} directly
linked to a quantum design, the quantities $A_1^A, A_1^B$ and $A_2$  are also 
second moments of a measurement of the observables $\lambda_i$ in random 
bases. The proof follows from a slight extension of the arguments given in
 Ref.~\cite{Tran2016}, see Appendix B \cite{SM}. This opens another possibility for an experimental implementation besides making randomized Pauli measurements on all the qubits individually. Now, we can formulate: 

{\bf Observation 3.} 
{\it 
Any two-qudit separable state obeys the relation
\begin{align}  
\label{obs3-twoqudits-threedim}
A_2 \leq d-1 + {(d-1)}A_1^A-A_1^B,
\end{align}
as well as the analogous one with parties $A$ and $B$ exchanged. This is equivalent 
to the criterion $S_2(\varrho_X) \leq  S_2(\varrho_{AB})$ for $X\in\{A,B\}$. This 
criterion is optimal, in the sense that if the inequality holds for $A_1^A, A_1^B$ 
and $A_2$, then there is a separable state compatible with these values.}

The criterion itself was established before, so we only 
have to prove the optimality statement. This is done in Appendix A \cite{SM}, where we explicitly 
construct the polytope of all  admissible values of $A_1^A, A_1^B$ and $A_2$ for general 
and separable states in any dimension. The unfortunate consequence of the optimality statement 
is that any PPT entanglement cannot be detected by the quantities $A_1^A, A_1^B$ and $A_2$ as the entropic criterion is strictly weaker than the PPT criterion \cite{Hiroshima2003}. In the following, we will overcome this obstacle by developing a general criterion for entanglement using higher moments of randomized 
measurements. 

{\it Higher-dimensional systems.---}
In higher-dimensional systems, different forms of entanglement exist 
e.g., entanglement of different dimensionality \cite{Terhal2000, Kraft2018} 
or bound entanglement \cite{MHorodecki1998, KHorodecki2005, Vertesi2014, TMoroder2014}. 
The previously known criteria for randomized measurements face serious
problems in this scenario. First, criteria using purities, such as Observation 3,
can only characterize states that violate the PPT criterion and
hence miss the bound entanglement. Second, while the notion of 
randomized  measurements as defined in
Eqs.~(\ref{eq-onemeasurement}, \ref{eq-moments1}) is independent of the dimension,
many results for qubits employ the concept of a Bloch sphere, which is not available 
for higher dimensions, where not all observables are equivalent
under randomized unitaries. 
Ref.~\cite{Tran2016} showed that some results for qubits are also
valid for higher dimensions as long as only second moments are 
considered, but these connections are definitely not valid 
for higher moments. 

To overcome these problems, we first note that a general observable is
characterized by its eigenvectors, determining the 
probabilities of the outcomes, and the eigenvalues, corresponding 
to the observed values. For computing the moments as in Eq.~(\ref{eq-moments1}),
the eigenvectors do not matter due to 
the averaging over all unitaries. 
The eigenvalues are relevant, but they may be altered in classical 
postprocessing: Once the frequencies of the outcomes are recorded, one 
can calculate the moments in Eq.~(\ref{eq-moments1}) for different
assignments of values to the outcomes.

So the question arises, whether one can choose the eigenvalues of 
an observable in a way, such that the moments in Eq.~(\ref{eq-moments1}) 
are easily tractable. For instance, it would be desirable to write
them as averages over a high-dimensional sphere (the so-called pseudo-Bloch
sphere). The reason is that several entanglement criteria, such as the 
computable cross norm or realignment criterion \cite{Rudolph2005, Chen2003} 
and the de Vicente (dV) criterion \cite{deVicente2007}, make also use of a 
pseudo Bloch sphere \cite{Tothinpreparation}. Surprisingly, 
the desired eigenvalues can always be found:

{\bf Observation 4.}
{\it
Consider an arbitrary observable in a higher-dimensional system.
Then, one can change its eigenvalues such that for the resulting observable $M_d$
the second and fourth moments $\mathcal{R}^{(r)}_{AB}$ 
in the sense of Eq.~(\ref{eq-moments1}) equal, up to a factor,
a moment $\mathcal{S}^{(r)}_{AB}$ which is taken by an integral 
over a generalized pseudo Bloch sphere. That is, $\mathcal{S}^{(r)}_{AB}$
is given by 
\begin{align} 
\label{eq-newmoments}
   \mathcal{S}^{(r)}_{AB} = N \int d\bm{\alpha}_1 \int d\bm{\alpha}_2
     [\mathrm{tr}
    (\varrho_{AB} \bm{\alpha}_1 \cdot \bm{\lambda} \otimes \bm{\alpha}_2 \cdot \bm{\lambda} )
    ]^r,
\end{align}
where $\bm{\alpha}_i$ denote $(d^2-1)$-dimensional unit real vectors
uniformly distributed from the pseudo Bloch sphere,
and $\bm{\lambda} = (\lambda_1, \lambda_2, \ldots, \lambda_{d^2-1} )$ is the
vector of Gell-Mann matrices. Furthermore, $N$ is a normalization factor.
}

The proof and the detailed form of $M_d$ are given in Appendix B \cite{SM}. To give a
simple example, for $d=3$ one may measure the standard spin measurement $J_z$ 
and assign the values $\alpha_+/\gamma$, $\alpha_-/\gamma$
and $2\beta/\gamma$ instead of the standard values $\pm 1$ and $0$ to
the three possible outcomes,
where
$\alpha_\pm = \pm 3 -\beta$, $\beta = -\sqrt{7 + 2 \sqrt{15}}$,
and $\gamma ={2\sqrt{5 + \sqrt{15}}}.$
Note that the resulting observable is also traceless.

\begin{figure}[t!]
    \centering
    \includegraphics[width=1.0\columnwidth]{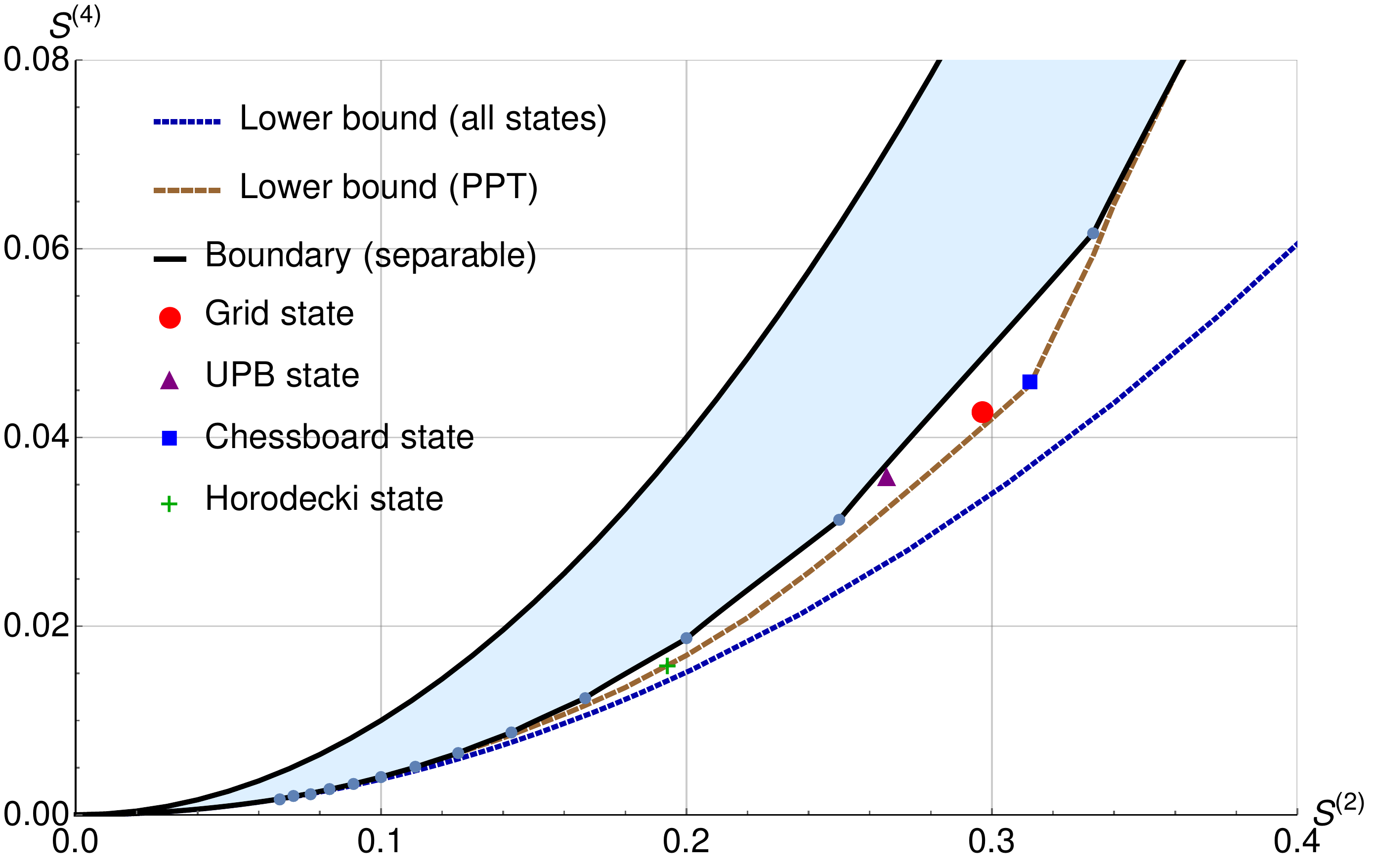}
    \caption{Entanglement criterion based on second and fourth 
    moments of randomized measurements for $3\otimes 3$ systems.
    Separable states are contained in the light-blue area,
    according to the discussion in the main text. Several bound
    entangled states (denoted by colored symbols) are outside,
    meaning that their entanglement can be detected with the methods
    developed in this paper. For comparison, we also indicate a
    lower bound on the fourth moment for PPT states, obtained by
    numerical optimization, as well as a bound for general states.
    Further details, such as the form of the states, are given in Appendix C. Information on the numerical methods is found in Appendix~D \cite{SM}.}
    \label{3by3figure}
\end{figure}

It remains to formulate separability criteria in terms of the second and fourth 
moments $\mathcal{S}^{(r)}_{AB}$. For that, we employ the dV criterion \cite{deVicente2007},
details of the calculations are given in Appendix B \cite{SM}. From these results, it also follows that the dV criterion can be evaluated via randomized measurements for all dimensions. First, it turns out that 
$\mathcal{S}^{(2)}_{AB}$ and $\mathcal{S}^{(4)}_{AB}$ can, for any dimension, be 
simply expressed as polynomial functions of the subset of the two-body correlation 
coefficients $t_{ij}$ with $1 \leq i,j \leq d^2-1$ in Eq.~(\ref{eq-twoqudits}), where
we also call this submatrix $T_{\mathrm s}$. Second, the moments $\mathcal{S}^{(r)}_{AB}$
are by definition invariant under orthogonal transformations of the matrix $T_{\mathrm s}$.
On the other hand, the dV criterion reads that two-qudit separable states obey
$\|T_{\mathrm s}\|_{\text{tr}}  \leq d-1$, and this is also invariant under the named 
orthogonal transformations. Third, for a fixed value of the second moment 
$\mathcal{S}^{(2)}_{AB}$, we can maximize and minimize the fourth moment  
$\mathcal{S}^{(4)}_{AB}$ under the constraint $\|T_{\mathrm s}\|_{\text{tr}}  \leq d-1$. 
This task is greatly simplified by orthogonal invariance; in fact, we can assume 
$T_{\mathrm s}$ to be diagonal. This leads to simple, piece-wise algebraic separability conditions for arbitrary dimensions $d$.

The results for $d=3$ are shown in Fig.~\ref{3by3figure}. The outlined procedure gives 
an area that contains all values of $\mathcal{S}^{(2)}_{AB}$ and $\mathcal{S}^{(4)}_{AB}$ 
for separable states. Most importantly, various bound entangled states can be detected \cite{Lockhart2018, Bruss2000, Bennett1999, PHorodecki1999}.
Also for $d=4$ bound entanglement can be detected, details are given in Appendix C \cite{SM}.

{\it Conclusion.---}
We have developed methods for characterizing quantum correlations using
randomized measurements. On the one hand, our approach led to optimal
criteria for different forms of entanglement using the second moments
of the randomized measurements. On the other hand, we have shown that
using fourth moments of randomized measurement detection of bound 
entanglement as a weak form of entanglement is possible. This opens
a new perspective for developing the approach further, as all 
previous entanglement criteria were only suited for highly entangled 
states. 

There are several directions for further research. First, on a more
technical level, the employed separability criterion \cite{deVicente2007}
can be derived from an approach towards entanglement using covariance
matrices \cite{Gittsovich2008}. Connecting randomized measurements to this approach
will automatically lead to further results, e.g., on the quantification of
entanglement \cite{Gittsovich2010}. Second, for experimental studies of the criteria
presented in this article, a scheme for the statistical analysis of finite
data, e.g., using the Hoeffding inequality or other large deviation bounds, 
is needed. Finally, our results encourage to develop the characterization of other quantum properties using randomized measurements, such as spin squeezing or the quantum Fisher information in metrology.

\begin{acknowledgments}
{\it Acknowledgments.---}
We thank Dagmar Bru{\ss} and Martin Kliesch for discussions. 
This work was supported by the Deutsche Forschungsgemeinschaft 
(DFG, German Research Foundation, project numbers 447948357 and 
440958198), the Sino-German Center for Research Promotion, and 
the ERC (Consolidator Grant No. 683107/TempoQ). N. W. acknowledges 
support by the QuantERA project QuICHE via the German Ministry
of Education and Research (BMBF Grant No. 16KIS1119K).
A. K. acknowledges support by the Georg H. Endress foundation.
\end{acknowledgments}

\vspace{0.5cm}
\onecolumngrid

\section*{Appendix A: Separability criteria with sector lengths}
In Appendix A, we first explain the notion of sector lengths
and known simple entanglement criteria. Second, we present a criterion 
of full separability and prove thereby Observation 1. Third, we propose a criterion 
of biseparability for fixed bipartitions and prove Observation 2.
Fourth, we discuss whether Observation 2 is also valid for mixtures of 
biseparable states with respect to different bipartitions. We discuss
numerical evidence for this conjecture and also present a proof for
a special case. Fifth, we discuss the improvements of our three-qubit 
separability criteria, in comparison with existing criteria. Sixth, we give 
a detailed discussion of Observation 3 and provide a proof. Finally, we 
characterize two-qudit states with sector lengths and discuss the strength
of the separability criterion.

\subsection*{A1. Sector lengths}
Let $\varrho$ be a $n$-particle and $d$-dimensional quantum
($n$-qudit) state.
The state $\varrho$ can be written in the generalized Bloch representation
\begin{align} \label{eq-nqudits}
\varrho = \frac{1}{d^{n}} \sum_{i_{1}, \cdots, i_{n}=0}^{d^2-1} \alpha_{i_{1} \cdots i_{n}}
\lambda_{i_{1}} \otimes \cdots \otimes \lambda_{i_{n}}, \tag{S.1}
\end{align}
where $\lambda_{0}$ is the identity, and 
$\lambda_{i}$
are the Gell-Mann matrices, normalized such that
$\lambda_i=\lambda_i^\dagger$,
$\mathrm{tr}\left[\lambda_{i} \lambda_{j}\right]=d \delta_{i j}$,
and $\mathrm{tr}\left[\lambda_{i} \right]=0$ for $i > 0$.
The real coefficients $\alpha_{i_{1} \cdots i_{n}}$ are given by
$\alpha_{i_{1} \cdots i_{n}}=\mathrm{tr}\left[\varrho \lambda_{i_{1}} \otimes \cdots \otimes \lambda_{i_{n}} \right]=\Braket{\lambda_{i_{1}} \otimes \cdots \otimes \lambda_{i_{n}}}$.
The state $\varrho$ can be represented by
\begin{align}
    \varrho = \frac{1}{d^{n}} \left(
    \eins^{\otimes n} + P_1 + P_2 + \cdots + P_n
    \right), \tag{S.2}
\end{align}
where the Hermitian operators $P_k$
for $k=1,2,\ldots,n$ denote the sum of all terms
coming from the basis elements with weight $k$
\begin{align}
    P_k(\varrho)= \sum_{\substack{i_{1}, \cdots, i_{n}=0,\\ \mathrm{wt}(\lambda_{i_{1}} \otimes \cdots \otimes \lambda_{i_{n}})=k }}^{d^2-1}
    \alpha_{i_{1} \cdots i_{n}}
\lambda_{i_{1}} \otimes \cdots \otimes \lambda_{i_{n}},\tag{S.3}
\end{align}
where the weight $\mathrm{wt}(\lambda_{i_{1}} \otimes \cdots \otimes \lambda_{i_{n}})$
is the number of non-identity Gell-Mann matrices.
Now, we can define sector lengths as
\begin{align}  \label{eq-sectorlengths}
    A_k(\varrho) =\frac{1}{d^n}\mathrm{tr}\left[P_k(\varrho)^2
    \right]
    = \sum_{\substack{i_{1}, \cdots, i_{n}=0,\\ \mathrm{wt}(\lambda_{i_{1}} \otimes \cdots \otimes \lambda_{i_{n}})=k }}^{d^2-1}
    \alpha_{i_{1} \cdots i_{n}}^2. \tag{S.4}
\end{align}
Physically, the sector lengths $A_k$ quantify the amount
of $k$-body quantum correlations. Note that $A_0=\alpha_{0 \cdots 0}=1$
due to $\text{tr}(\varrho)=1$. The sector lengths $A_k$ can be associated
with the purity
of $\varrho$:
\begin{align} \label{eq-sector-purity}
    \mathrm{tr}(\varrho^2) = \frac{1}{d^{n}} \sum_{i_{1}, \cdots, i_{n}=0}^{d^2-1} \alpha_{i_{1} \cdots i_{n}}^2 = \frac{1}{d^{n}} \sum_{k=0}^{n}A_k(\varrho). \tag{S.5}
\end{align}
As the simplest case, if we consider a single-qubit state ($n=1$ and $d=2$),
it holds that $\mathrm{tr}(\varrho^2)=(1+A_1)/2 \leq 1$, which
leads to $A_1 \leq 1$.

There are three useful properties of sector lengths. The first one is local-unitary 
invariance,
$A_k(\varrho)=A_k\big(U_{1} \otimes \cdots \otimes U_{n} \varrho U_{1}^{\dagger} \otimes \cdots \otimes U_{n}^{\dagger}\big)$.
The second one is convexity,
$A_k \left(\sum_i p_i \ket{\psi_i}\!\bra{\psi_i}\right) \leq \sum_i p_i A_k(\ket{\psi_i})$.
The third one is a convolution property:
for a $n$-particle state $\varrho_P \otimes \varrho_Q$,
$A_k(\varrho_P \otimes \varrho_Q) = \sum_{i=0}^{k} A_i(\varrho_P) A_{k-i}(\varrho_Q)$,
where $\varrho_P$ and $\varrho_Q$ are respectively $j$-particle
and $(n-j)$-particle states. Due to these properties, the sector lengths
can be useful for entanglement detection
\cite{Aschauer2003, Vicente2011, Klockl2015, Nikolai2020}.

For example, for all two-qubit separable states
($n=d=2$), it holds that
\begin{align} 
       A_2\big(\sum_i p_i
       \ket{\psi_i^A}\!\bra{\psi_i^A}
       \otimes
       \ket{\psi_i^B}\!\bra{\psi_i^B}
       \big) 
  \leq \sum_i p_i A_2 \left(\ket{\psi_i^A}\!\bra{\psi_i^A}
       \otimes
       \ket{\psi_i^B}\!\bra{\psi_i^B} \right)
  =    \sum_i p_i A_1 (\ket{\psi_i^A}) 
  A_1 (\ket{\psi_i^B}) 
  \leq \sum_i p_i = 1. \tag{S.6}
\end{align}
Conversely, if $A_2>1$, then the two-qubit state is entangled.
In fact, since the Bell state $\ket{{\Phi^+}}=(\ket{00}+\ket{11})/\sqrt{2}$
can be written as
$
\ket{{\Phi^+}}\!\bra{{\Phi^+}} =
\left(\mathbb{I} \otimes \mathbb{I}
+ X \otimes X
- Y \otimes Y
+ Z \otimes Z
\right)/4
$,
its sector lengths are given by $(A_1,A_2)=(0,3)$, violating the separability criterion $A_2 \leq 1$.

\subsection*{A2. Criterion of full separability and proof of Observation 1}
Now we present the three-qudit generalization of the
fully separability criterion (\ref{obs1-fs}) in Observation 1 
in the main text.

\noindent
{\bf Observation 5.} 
{
\it Any fully separable 
three-qudit state obeys
\begin{align}  \label{obs5-fs-qudits}
    A_3 \leq d-1 + \frac{2d-3}{3}A_1 + \frac{d-3}{3} A_2. \tag{S.7}
\end{align}
}

\begin{remark}
In $d=2$, one obtains
Eq.~(\ref{obs1-fs}): $A_2 + 3A_3 \leq 3 +A_1$,
also shown in Fig.\,\ref{threequbits} in the main text.
Since one can easily construct fully separable states on two of the three sides of 
the resulting triangle (i.e., the surface where equality holds), this criterion is 
optimal in the sense that any other linear criterion for the $A_k$ detects strictly 
fewer states. Extensive numerical search suggests, however, that there are points 
on the triangle surface plane which cannot originate from a separable state. This may
indicate that there exist stronger, non-linear criteria for full separability  using 
sector lengths. For more details on the numerical optimization, see Appendix~D.
\end{remark}

\begin{proof}
First, recall that a three-particle state is fully separable
if it can be written as
\be 
\varrho_{\text{fs}} = \sum_k p_k \vr^A_k \otimes \vr^B_k \otimes \vr^C_k. \tag{S.8}
\ee
Let us consider the reduced density matrix on
the subsystem $AB$:
$\varrho_{AB} = {\mathrm{tr}}_{C}(\varrho_{\text{fs}}) = \sum_k p_k \vr^A_k \otimes \vr^B_k$.
Then we have
\begin{align} 
    {\mathrm{tr}} (\varrho_{\text{fs}}^2)
    = \sum_{k,l} p_k p_l \, \mathrm{tr}(\vr^A_k \vr^A_l) \ \mathrm{tr}(\vr^B_k \vr^B_l) \ \mathrm{tr}(\vr^C_k \vr^C_l)
    \leq
    \sum_{k,l} p_k p_l \, \mathrm{tr}(\vr^A_k \vr^A_l) \ \mathrm{tr}(\vr^B_k \vr^B_l)
    = {\mathrm{tr}} (\varrho_{AB}^2), \tag{S.9}
\end{align}
where $\mathrm{tr}(\vr^C_k \vr^C_l) \leq \sqrt{\mathrm{tr}(\vr^C_k)^2}\sqrt{\mathrm{tr}(\vr^C_l)^2} \leq 1$. Similarly, we obtain
${\mathrm{tr}} (\varrho_{\text{fs}}^2)  \leq  {\mathrm{tr}} (\varrho_{AC}^2)$ and
${\mathrm{tr}} (\varrho_{\text{fs}}^2)  \leq  {\mathrm{tr}} (\varrho_{BC}^2)$.
Summarizing these three purity inequalities gives
\begin{align} 
    3{\mathrm{tr}} (\varrho_{\text{fs}}^2)  \leq  {\mathrm{tr}} (\varrho_{AB}^2) +
    {\mathrm{tr}} (\varrho_{AC}^2) + {\mathrm{tr}} (\varrho_{BC}^2). \tag{S.10}
\end{align}
Then, with the help of the relation (\ref{eq-sector-purity}), translating this
inequality to the form with sector lengths yields Eq.~(\ref{obs5-fs-qudits}).
\end{proof}

\subsection*{A3. Criterion of biseparability and proof of Observation 2}
Now we discuss the three-qudit generalization of the
biseparability criterion (\ref{obs2-bs}) in Observation 2 
in the main text.

\noindent
{\bf Observation 6.} 
{\it Any three-qudit state which is separable
with respect to
some bipartition obeys
\begin{align}  \label{obs6-bs-qudits}
   A_2+A_3\leq \frac{d^3-2}{2}(1+A_1). \tag{S.11}
\end{align}
}

\begin{figure}[t]
    \centering
    \includegraphics[width=0.7\columnwidth]{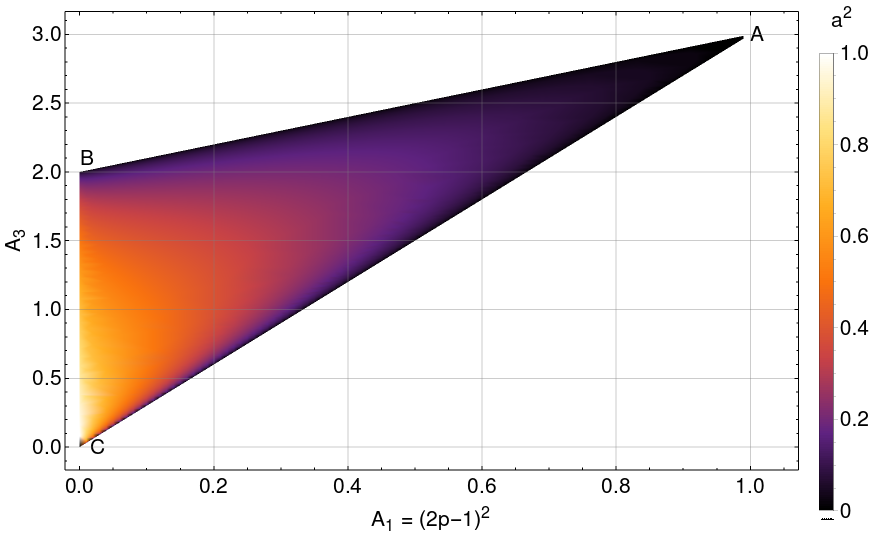}
    \caption{Location of states from the family defined in Eq.~(\ref{eq:bisep_family}) covering the plane given by Eq.~(\ref{obs2-bs}) in a $(A_1$,$A_3$)-plot. While the parameter $p$ fixes the $A_1$-coordinate via $A_1 = (2p-1)^2$, the parameter $a$ together with a choice of sign for $d$ defines the $A_3$-coordinate. The labelled vertices correspond to the states $\ketbra{0} \otimes \ketbra{\Phi^+}$ (A), $\frac12\ketbra{0}\otimes \ketbra{\Phi^+}+\frac12\ketbra{1}\otimes \ketbra{\Phi^-}$ (B) and $\frac12\ketbra{0}\otimes \ketbra{\Phi^+}+\frac12\ketbra{1}\otimes \ketbra{\Phi^+}$ (C).
    }
    \label{fig:bisep_family}
\end{figure}

\begin{remark}
In $d=2$, one obtains
Eq.~(\ref{obs2-bs}): $A_2+A_3 \leq 3 (1+A_1)$,
as shown in Fig.\,\ref{threequbits} in the main text.
This is the optimal criterion in the sense that
if the three $A_k$ obey the inequality, then for any bipartition
there is a separable state compatible with them.
This can be seen as follows: Eq.~(\ref{obs2-bs}) 
is saturated by a family of biseparable three-qubit states
\begin{align} \label{eq:bisep_family}
        \sigma= p\ket{0}\!\bra{0} \otimes
        \ket{\psi}\!\bra{\psi} + (1-p) \ket{1}\!\bra{1} \otimes
        \ket{\phi}\!\bra{\phi}, \tag{S.12}
\end{align}
    where
    \begin{align}  \nonumber
    &\ket{\psi} = a \ket{00} + b \ket{11},
    \ \ \ \ket{\phi} = c \ket{00} + d \ket{11},
    \ \ \ a^2+b^2 = c^2+d^2 =1,
    \ \ \ \frac{1}{2} \leq p \leq 1,
    \\
    &\sqrt{1-\frac{1}{2p}} \leq a \leq \sqrt{\frac{1}{2p}},
    \ \ \ a^2 \leq b^2,
    \ \ \ c = \sqrt{\frac{2pa^2-1}{2(p-1)}},
    \ \ \ d=\pm \sqrt{1-c^2}. \tag{S.13}
\end{align}
Since the state $\sigma$ has the sector lengths
\begin{align}
    A_1&= (2 p - 1)^2, \tag{S.14}\\ 
    A_2&= 1 + 2[2pab + 2(1 - p)cd]^2 +  2[p(a^2 - b^2) - (1 - p)(c^2 - d^2)]^2, \tag{S.15}\\ 
    A_3&= (2 p - 1)^2 + 2[2 pab- 2(1 - p)cd]^2, \tag{S.16}
\end{align}
it satisfies the equality $A_2+A_3 = 3(1+A_1)$ and lies on the plane displayed
as the boundary between the red and green areas in the polytope
in Fig.\,\ref{threequbits}. In order to see that this family indeed fills the entired 
plane, we display the plane and states from the family in Fig.~(\ref{fig:bisep_family}).
\end{remark}

\begin{proof}
Let $\varrho_{\text{bs}}$ be a separable three-qudit state
with respect to a bipartition, and let $\varrho_{X}$ be its
reduced density matrices on the subsystems $X$
for $X \in \{A,B,C\}$.
With the help of the relation (\ref{eq-sector-purity}), 
we notice that Eq.~(\ref{obs6-bs-qudits}) is equivalent to
\begin{align}
    1 +\mathrm{tr}(\varrho_{\text{bs}}^2)
    \leq \frac{d}{2} \left[
    \mathrm{tr}(\varrho_{A}^2) +
    \mathrm{tr}(\varrho_{B}^2) +
    \mathrm{tr}(\varrho_{C}^2)
    \right]. \tag{S.17}
\end{align}
In the following, without loss of  generality, we consider
a separable state with respect to a fixed
bipartition $A|BC$:
\be 
\varrho_{{A|BC}} = \sum_k q_k \vr^A_k \otimes \vr^{BC}_k. \tag{S.18}
\ee
Since its reduced density matrix on the
subsystem $A$ is given by
$\varrho_{{A}} = \sum_k q_k \vr^A_k$,
we have
\begin{align} 
    {\mathrm{tr}} (\varrho_{{A|BC}}^2)
    = \sum_{k,l} q_k q_l \, \mathrm{tr}(\vr^A_k \vr^A_l) \ \mathrm{tr}(\vr^{BC}_k \vr^{BC}_l)
    \leq \sum_{k,l} q_k q_l \, \mathrm{tr}(\vr^A_k \vr^A_l)
    = {\mathrm{tr}} (\varrho_{{A}}^2), \tag{S.19}
\end{align}
where $\mathrm{tr}(\vr^{BC}_k \vr^{BC}_l) \leq 1$.
In addition, it follows from the relation (\ref{eq-sector-purity}) for single particles
\begin{align} 
    \mathrm{tr}(\varrho_{B}^2) + \mathrm{tr}(\varrho_{C}^2) = \frac{1}{d} (2+A_1^B+A_1^C), \tag{S.20}
\end{align}
where we split $A_1=A_1^A+A_1^B+A_1^C$
corresponding to the contributions from the three particles.
Then we have
\begin{align} \nonumber
    1 + {\mathrm{tr}} (\varrho_{{A|BC}}^2)
    - \frac{d}{2} \left[
    \mathrm{tr}(\varrho_{A}^2) +
    \mathrm{tr}(\varrho_{B}^2) +
    \mathrm{tr}(\varrho_{C}^2)
    \right] 
    &\leq
    1 + \mathrm{tr}(\varrho_{A}^2)
    - \frac{d}{2} \left[
    \mathrm{tr}(\varrho_{A}^2) +
    \mathrm{tr}(\varrho_{B}^2) +
    \mathrm{tr}(\varrho_{C}^2)
    \right] \\ \nonumber
    &=
    1 + \left(1-\frac{d}{2}\right) {\mathrm{tr}} (\varrho_{{A}}^2)
    - \frac{1}{2} (2+A_1^B+A_1^C) \\ \nonumber
    &=
    \left(1-\frac{d}{2}\right) {\mathrm{tr}} (\varrho_{{A}}^2)
    - \frac{1}{2} (A_1^B+A_1^C) \leq 0.
\end{align}
\end{proof}

\subsection*{A4. Evidence for the validity of Observation 2 for mixtures of different bipartitions}

While Observation 6 holds true for mixed states of fixed bipartitions, we 
collected evidence for the conjecture that for $d=2$, the bound is also
true for mixtures of different bipartitions. To that end, we numerically
maximized the expression $A_2 + A_3 - 3(1+A_1)$ for mixtures of up to eight
biseparable three-qubit states with different bipartitions and found no violation (see Appendix~D for details). 
Furthermore, we can analytically show that mixtures of two product states for 
different bipartitions obey the bound as well:

\noindent
{\bf Observation 7.}
{\it
For three-qubit systems, a rank-$2$ mixture of two
pure biseparable states with respect to
different partitions obeys
the biseparability criterion (\ref{obs2-bs})
in the main text.
}

\begin{proof}
Let $\varrho$ be a rank-$2$ mixture of two
pure biseparable states with respect to
different partitions, without loss of generality,
$A|BC$ and $AB|C$:
\begin{align} 
    \varrho = p \ket{\Psi}\!\bra{\Psi}
    +(1-p)\ket{\Phi}\!\bra{\Phi}, \tag{S.21}
\end{align}
where $0\leq p \leq 1$, and without loss of generality, we can write
\begin{align} \nonumber
    \ket{\Psi}&=\ket{0}
    \otimes
    \left(
    \lambda_0 \ket{00} + \lambda_1 \ket{11}
    \right), \ \ \ \ \ \
    0 \leq \lambda_0, \, \lambda_1 \leq 1,
    \ \ \ \ \ \
    \lambda_0^2 + \lambda_1^2 =1,
    \\ 
    \ket{\Phi}&=
    \sum_{i,j=0}^1\kappa_{ij}\ket{ij}
     \otimes
    \left(c_0\ket{0}+c_1\ket{1}
    \right), \ \ \ \ \ \
    \kappa_{ij},\, c_1,\, c_2 \in \mathbb{C},
    \ \ \ \ \ \
    \sum_{i,j=0}^1|\kappa_{ij}|^2
    =|c_0|^2 + |c_1|^2 =1. \tag{S.22}
\end{align}
Now, we define a function
\begin{align} 
    f(\varrho_1, \varrho_2)
    \equiv
    1 +\mathrm{tr}(\varrho_1 \varrho_2)
    -\mathrm{tr}(\varrho_{A_1} \varrho_{A_2})
    -\mathrm{tr}(\varrho_{B_1} \varrho_{B_2})
    -\mathrm{tr}(\varrho_{C_1} \varrho_{C_2}), \tag{S.23}
\end{align}
where $\varrho_i$ $(i=1,2)$ are three-particle
quantum states and $\varrho_{X_i}$ for $X \in \{A,B,C\}$
are their reduced density matrices on the
subsystems $X$.
Our aim is to prove that the biseparable
state $\varrho$ obeys
\begin{align} 
    1 +\mathrm{tr}(\varrho^2)
    \leq 
    \mathrm{tr}(\varrho_{A}^2)
    +\mathrm{tr}(\varrho_{B}^2)
    +\mathrm{tr}(\varrho_{C}^2). \tag{S.24}
\end{align}
This is equivalent to proving that
the following function $F(\varrho)$ is non-positive:
\begin{align} \nonumber
    F(\varrho)
    &=
    1 +\mathrm{tr}(\varrho^2)
    -\mathrm{tr}(\varrho_{A}^2)
    -\mathrm{tr}(\varrho_{B}^2)
    -\mathrm{tr}(\varrho_{C}^2)
    \\ 
    &=p^2f(\Psi, \Psi)
    + (1-p)^2f(\Phi, \Phi)
    +2p(1-p)f(\Psi,\Phi). \tag{S.25}
\end{align}
A straightforward calculations yields
\begin{align} 
    f(\Psi, \Psi)&=1-2(\lambda_0^4 + \lambda_1^4), \tag{S.26}
    \\ 
    f(\Phi, \Phi)&=1-2\mathrm{tr}
    (\kappa \kappa^{\dagger} \kappa \kappa^{\dagger}), \tag{S.27}
    \\ 
    f(\Psi, \Phi)&=1
    +|\lambda_0 \kappa_{00} c_0
    +\lambda_1 \kappa_{01} c_1|^2
    -(\kappa_{00}^2 + \kappa_{01}^2)
    -\left[\lambda_0^2 (\kappa_{00}^2 + \kappa_{10}^2)
    + \lambda_1^2
    (\kappa_{01}^2 + \kappa_{11}^2)\right]
    -\left(\lambda_0^2 |c_0|^2 +
    \lambda_1^2 |c_1|^2
    \right), \tag{S.28}
\end{align}
where $\kappa=(\kappa_{ij})$.
Here, since $\kappa_{ij}$ and $c_i$ can be taken as
real, we obtain 
$|\lambda_0 \kappa_{00} c_0
    +\lambda_1 \kappa_{01} c_1|^2 \leq
    2\left(\lambda_0^2 \kappa_{00}^2 c_0^2
    +\lambda_1^2 \kappa_{01}^2 c_1^2
    \right)$.
Also, for the $2 \times 2$ matrix $\kappa$
we know that $1-2\mathrm{tr}(\kappa \kappa^{\dagger} \kappa \kappa^{\dagger}) = 4(\det \kappa)^2-1$.
Thus we have
\begin{align} \nonumber
    F(\varrho) &\leq
    p^2 \left[1-2(\lambda_0^4 + \lambda_1^4)\right]
    + (1-p)^2\left[4(\det \kappa)^2-1 \right]\\ 
    &\quad+2p(1-p)\left[
    1+\lambda_0^2 c_0^2 (2\kappa_{00}^2 -1)
    +\lambda_1^2 c_1^2 (2\kappa_{01}^2 -1)
    -(\kappa_{00}^2 +\kappa_{01}^2)
    -\lambda_0^2 (\kappa_{00}^2 + \kappa_{10}^2)
    -\lambda_1^2 (\kappa_{01}^2 +\kappa_{11}^2)
    \right]. \tag{S.29}
\end{align}
Now, it is sufficient to show that the maximization
of the right-hand side is non-positive.
In fact, the best choice is to set
\begin{align} 
    c_0^2&=1, \ \ \ \ c_1^2=0, \ \ \ \ \
    \mathrm{if} \ \
    (2\kappa_{01}^2 -1) \lambda_1^2 \leq
    (2\kappa_{00}^2 -1)\lambda_0^2, \tag{S.30}
    \\ 
    c_1^2&=1, \ \ \ \ c_0^2=0, \ \ \ \ \
    \mathrm{if} \ \
    (2\kappa_{00}^2 -1)\lambda_0^2 \leq
    (2\kappa_{01}^2 -1) \lambda_1^2. \tag{S.31}
\end{align}
Let us consider the former case: $c_0^2=1$ and $c_1^2=0$.
Due to that
$\sum_{i,j} \, \kappa_{ij}^2=\lambda_0^2 +\lambda_1^2 =1$, we find
\begin{align} 
    F(\varrho) \leq
    p^2 \left\{
    1-2\left[\lambda_0^4 + (1-\lambda_0^2)^2\right]
    \right\}
    + (1-p)^2\left[4(\det \kappa)^2-1 \right] +2p(1-p)\left(
    \kappa_{10}^2-\kappa_{01}^2
    -2\lambda_0^2 \kappa_{10}^2 \right). \tag{S.32}
\end{align}
Maximization of the right-hand side with
respect to $\lambda_0^2$ can be achieved by
three cases:
(1) $\lambda_0^2 = 0$, 
(2) $\lambda_0^2 = 1$, 
(3) $\lambda_0^2 = [1-(1-p)\kappa_{10}^2/2p]/2$ if $\kappa_{10} < {p}/({1-p})$.
In all cases, we can immediately
show that $F(\varrho) \leq 0$.
\end{proof}

\subsection*{A5. Discussion of the three-qubit separability criteria}
Here, we focus on the case of qubit systems.
The existing criteria are as follows.
(i) any fully separable three-qubit state
obeys $A_3\leq 1$ \cite{Vicente2011, Tran2016}.
If this inequality is violated, the state
is entangled but it may be still separable for some
bipartition. (ii) any biseparable
three-qubit state obeys $A_3\leq 3$ \cite{Vicente2011, Nikolai2020}.
If this is
violated, the state is genuinely tripartite
entangled.
Note that these existing criteria
can straightforwardly be derived from the
convexity and convolution
of sector lengths.
In the following, we will show that
our criteria (\ref{obs1-fs}, \ref{obs2-bs})
significantly improve the existing criteria,
introducing some examples.

\begin{figure}[t]
    \centering
    \includegraphics[width=0.35\columnwidth]{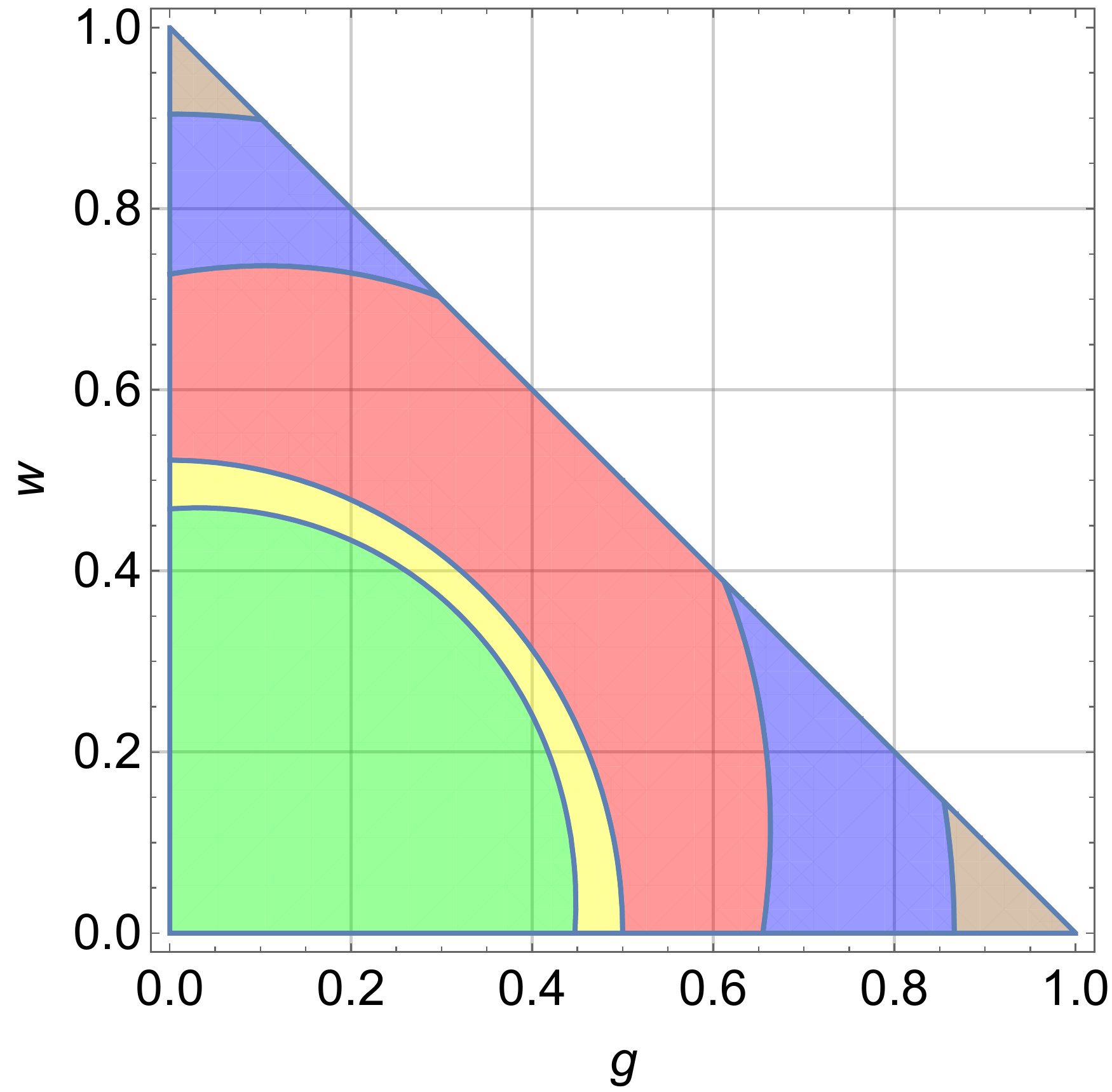}
    \caption{
    Entanglement criteria for the noisy GHZ-W state (\ref{noisyGHZWstate}) in the $g-w$ plane. Previously, several works
    \cite{Huber2010, Szalay2011} have discussed entanglement
    criteria in this two-parameter space.
    The fully separable states are contained in the green
    area, obeying our criterion (\ref{obs1-fs}).
    The outside of the green and yellow areas corresponds to
    the biseparable or genuine entangled states that violate
    a previously known criterion for fully separable states, $A_3 \leq 1$.
    Thus, the yellow area marks the improvement of Observation 1 compared with previous results.
    Also, states that are biseparable for some partitions
    are contained in the union of the green, yellow, and red areas, characterized by our criterion (\ref{obs2-bs}).
    The brown area corresponds to the genuine entangled states violating a previously known criterion for biseparable states, $A_3 \leq 3$.
    Thus, the blue area marks the improvement of Observation 2 compared with previous results.
    }
    \label{ghzwfigure}
\end{figure}

A good example for three-qubit states are the
noisy GHZ-W mixed states
\cite{Lohmayer2006, Szalay2011}:
\be \label{noisyGHZWstate}
\varrho =
g\ket{\mathrm{GHZ}}\bra{\mathrm{GHZ}}
+w\ket{\mathrm{W}}\bra{\mathrm{W}}
+\frac{1-g-w}{8}\eins^{\otimes 3}, \tag{S.33}
\ee
where $0\leq g,w\leq 1$, and the GHZ state and the W state are
given by
\be 
\ket{\mathrm{GHZ}} =\frac{1}{\sqrt{2}} \left(\ket{000} + \ket{111}\right),
\ \ \ \
\ket{\mathrm{W}} =\frac{1}{\sqrt{3}} \left(\ket{001} + \ket{010} + \ket{100}\right), \tag{S.34}
\ee
where 
$\Braket{\mathrm{GHZ}|\mathrm{W}}=0$.
The noisy GHZ-W mixed state has
$(A_1, A_2, A_3) = \left(w^2/3, 3g^2+3w^2-2gw, 4g^2+11w^2/3\right)$.
To analyze this state, we consider three
cases: (i) the noisy GHZ state, i.e.,  $w=0$
(ii) the noisy W state, i.e.,  $g=0$ (iii)
the GHZ-W mixed state, i.e., $g+w=1$.
Tables \ref{tab-1} and \ref{tab-2} list the
results of our criteria, comparing them to the
existing criteria and the optimal values.
Also, the criteria for the
state (\ref{noisyGHZWstate}) are
illustrated on the $g-w$ plane 
in Fig.~\ref{ghzwfigure}.

\begin{table}[]
    \centering
\begin{tabular}{ |p{4.5cm}|p{3cm}|p{3cm}|p{3cm}|}
\hline
\multicolumn{4}{|c|}{Criteria for full separability} \\
\hline
Three-qubit states&$A_3\leq 1$&Eq.~(\ref{obs1-fs})&Optimal values \\
\hline
$w=0$: noisy GHZ mixture&
$g\leq 0.5$&
$g\leq \frac{1}{\sqrt{5}} \approx 0.447$&
$g\leq  0.2$ \cite{Dur1999}\\
$g=0$: noisy W mixture&
$w\leq \sqrt{\frac{3}{11}}\approx 0.522$&
$w\leq \frac{3}{\sqrt{41}} \approx 0.469$&
$w\leq 0.177$ \cite{Chen2012}\\
$g+w=1$: GHZ-W mixture&
all states detected&
all states detected&
\\
\hline
\end{tabular}
    \caption{
Results for the fully separable criterion in Eq.~(\ref{obs1-fs})
in the main text,
compared with the existing criterion $A_3\leq 1$ and the optimal values.
For $w=0$ or $g=0$, the noisy mixed
GHZ and W state are known to be fully
separable iff $g\leq 0.2$ \cite{Dur1999}
and $w\leq 0.177$ \cite{Chen2012}.
Clearly, the bound (\ref{obs1-fs}) improves the existing
bound $A_3\leq 1$. 
}
\label{tab-1}
\end{table}

\begin{table}[]
    \centering
\begin{tabular}{ |p{4.5cm}|p{3cm}|p{3cm}|p{3cm}|}
\hline
\multicolumn{4}{|c|}{Criteria for biseparability} \\
\hline
Three-qubit states&$A_3\leq3$&Eq.~(\ref{obs2-bs})&Optimal values\\
\hline
$w=0$: noisy GHZ mixture&
$g\leq \frac{\sqrt{3}}{2} \approx 0.866$&
$g\leq \sqrt{\frac{3}{7}} \approx 0.655$&
$g\leq \frac{3}{7} \approx 0.429$
\cite{Guhne2010} \\
$g=0$: noisy W mixture&
$w\leq \frac{3}{\sqrt{11}} \approx 0.905$&
$w\leq \frac{3}{\sqrt{17}} \approx 0.728$&
$w\leq  0.479$ \cite{Jungnitsch2011}\\
$g+w=1$: GHZ-W mixture&
$0.102 \leq g \leq 0.855$&
$0.297 \leq g \leq 0.612 $&
 \\
\hline
\end{tabular}
    \caption{
Results for the biseparable criterion Eq.~(\ref{obs2-bs})
in the main text, compared with the existing criterion $A_3\leq 3$ 
and the optimal values.
For $w=0$ or $g=0$, the noisy mixed GHZ and W state are known to be
biseparable iff $g\leq 0.429$ \cite{Guhne2010} and $w\leq  0.479$ \cite{Jungnitsch2011}.
For $g+w=1$ ($w=1-g$), the existing
criterion and our criterion (\ref{obs2-bs}), respectively,
imply that the GHZ-W mixed state can be
biseparable only in some interval for $g$.
Interestingly, Ref.~\cite{Lohmayer2006} has analyzed
the GHZ-W mixed states using the three-tangle $\tau$ 
and the squared concurrences $C_{XY}^2$ measuring 
bipartite entanglement in the reduced states (note that all 
reduced states are equivalent).
It has been shown that 
for a region
$0.292 \leq g \leq 0.627$, the 
state has zero three-tangle and zero
concurrence in the reduced states .
This region is larger than the region which is not detected by
Eq.~(\ref{obs2-bs}).
Thus, Observation 2 can detect
multiparticle entanglement 
even when the three-tangle as well as bipartite entanglement
in reduced states vanishes.
}
\label{tab-2}
\end{table}

\subsection*{A6. Criterion of separability and proof of Observation 3}
Let us present the more general description of
the separability criterion (\ref{obs3-twoqudits-threedim})
in Observation 3 in the main text.

\noindent
{\bf Observation 8.}
{\it
Any two-qudit separable state obeys the relation
\begin{align}  \label{obs3-twoquditscriterion}
 A_2 &\leq d-1 + \min \{ (d-1)A_1^A -A_1^B, (d-1)A_1^B -A_1^A\}. \tag{S.35}
\end{align}
For $A_1^A \leq A_1^B$, this relation becomes
\begin{align} \label{ob3-A1A<A1Bcase}
A_2 \leq d-1 + {(d-1)}A_1^A-A_1^B, \tag{S.36}
\end{align}
as well as the analogous one with parties $A$ and $B$ exchanged.
This is equivalent to the criterion
$S_2(\varrho_X) \leq S_2(\varrho_{AB})$ for $X\in\{A,B\}$,
where $S_2$ denotes the second-order R\'{e}nyi entropy
and $\varrho_X$ denote the reduced density matrices.
This criterion is optimal, in the sense that if the inequality holds
for $A_1^A, A_1^B$ and $A_2$, then
there is a separable state compatible with these values.
}

\begin{figure}
    \centering
    \includegraphics[width=0.5\columnwidth]{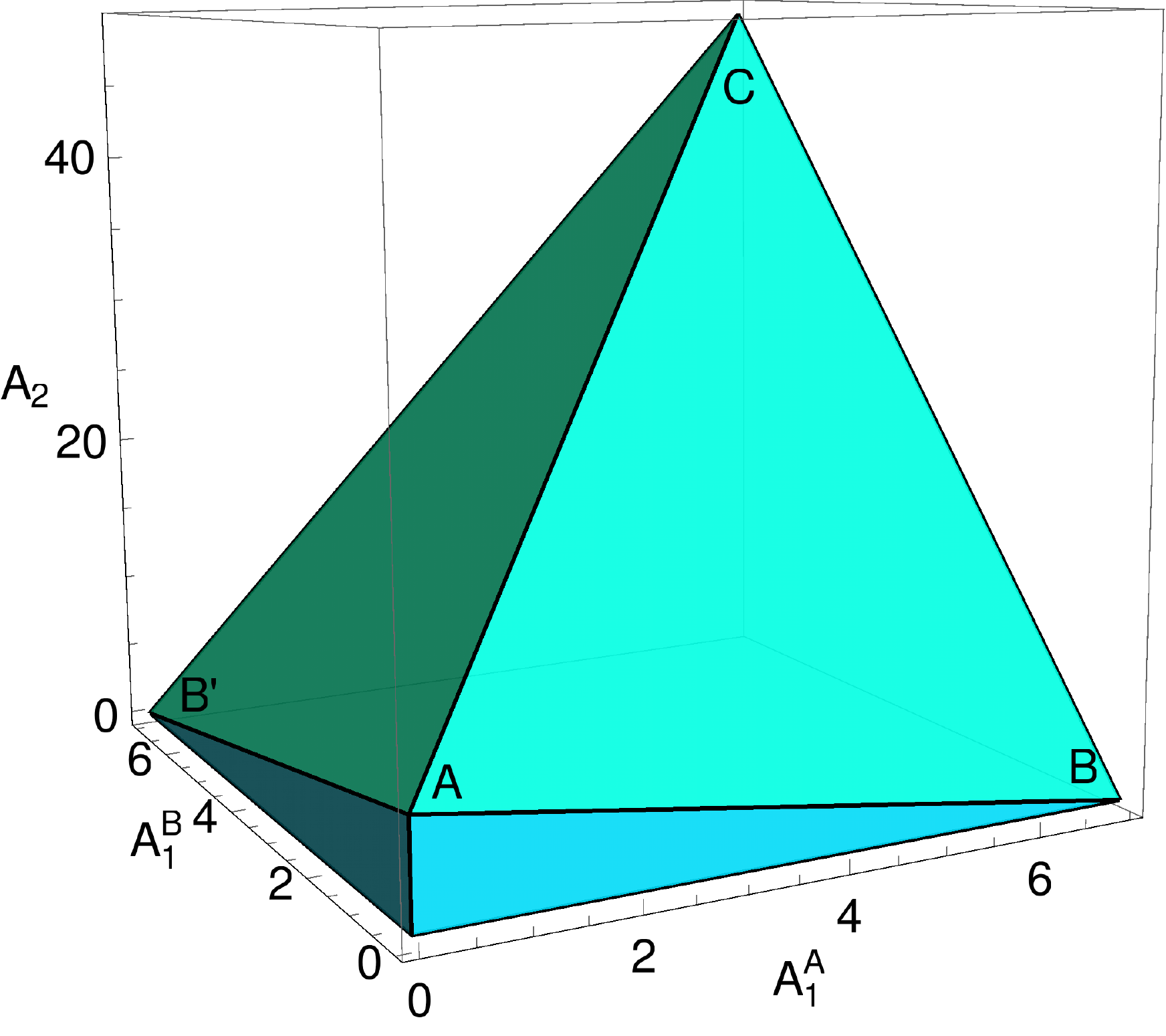}
    \caption{The polytope of bipartite separable states of $d \otimes d$-dimensional systems in terms of the quantities $A_1^A, A_1^B,$
    and $A_2$, where $d=8$.
    The two triangular surfaces on the top originate from the constraints in Eq.~(\ref{ob7-prf-2}) and are covered by states $\varrho(p, \theta)$ from the family in Eq.~(\ref{obs7-twoqudits}). The vertices of the triangle correspond to parameters $p=1/d$, $\theta={\pi}/2$ (A), $p=1/d$, $\theta=0$ (B') and $p=1$ (C). Point (B) is obtained by exchanging the role of A and B in $\varrho(1/d,0)$.
    Note that the construction of the surface also requires the knowledge of
    the entire polytope, this is derived in Section A7.}
    \label{fig:bipartite_family}
\end{figure}

\begin{proof}
Let $\varrho_{\text{sep}}$ be a two-qudit separable state.
Here, the entropic criterion
\cite{HorodeckiMajori1996, Elben2018}
states that any bipartite separable state obeys
that 
$S_2(\varrho_A) \leq S_2(\varrho_{\text{sep}})$
and $S_2(\varrho_B) \leq S_2(\varrho_{\text{sep}})$,
where $\varrho_X$ denote the reduced density matrices of $\varrho_{\text{sep}}$.
The entropic inequalities can be written as
\begin{align} \label{ob7-prf-1}
    \mathrm{tr}(\varrho_{\text{sep}}^2) \leq \mathrm{tr}(\varrho_A^2),
    \ \ \ \ \
    \mathrm{tr}(\varrho_{\text{sep}}^2) \leq \mathrm{tr}(\varrho_B^2). \tag{S.37}
\end{align}
Using the relation (\ref{eq-sector-purity}), we can
respectively translate these inequalities to
\begin{align} \label{ob7-prf-2}
    A_2 \leq d-1 + (d-1)A_1^A -A_1^B,
    \ \ \ \ \  
    A_2 \leq d-1 + (d-1)A_1^B -A_1^A. \tag{S.38}
\end{align}
Thus we have Eq.~(\ref{obs3-twoquditscriterion}).

The novel point is proving the optimality.
In the following, we show that for $A_1^A \leq A_1^B$, Eq.~(\ref{ob3-A1A<A1Bcase}) is
saturated by a family of separable states
\begin{align} \label{obs7-twoqudits}
       \varrho(p,\theta) = p\ket{00}\!\bra{00} + q \sum_{j=1}^{d-1} \ket{j}\!\bra{j}\otimes \ket{\theta_{0j}}\!\bra{\theta_{0j}}, \tag{S.39}
\end{align}
where $\ket{\theta_{ij}} = \cos{\theta}\ket{i} + \sin{\theta}\ket{j}$,
$q={(1-p)}/({d-1})$, $p \in [1/d,1]$, and $\theta \in [0,{\pi}/2]$.
Note that a family for the other case ($A_1^B \leq A_1^A$) can be found if
the two parties of $\varrho(p,\theta)$ are interchanged.
In fact, from Eqs.~(\ref{ob7-prf-1}, \ref{ob7-prf-2}), we
immediately notice that Eq.~(\ref{ob3-A1A<A1Bcase})
is saturated iff $\mathrm{tr}(\varrho_{\text{sep}}^2) = \mathrm{tr}(\varrho_A^2)$.
For the state (\ref{obs7-twoqudits}), we find 
\begin{align} 
\mathrm{tr}(\varrho_A^2) &= p^2 + (d-1)q^2 = \mathrm{tr}
\left(
\varrho(p,\theta)^2
\right), \tag{S.40} \\ 
\mathrm{tr}(\varrho_B^2) &= \mathrm{tr}(\varrho_A^2) + 2(d-1)pq\cos^2(\theta) + 2\binom{d-1}{2}q^2\cos^4(\theta), \tag{S.41}
\end{align}
which results in
\begin{align} 
    A_1^A &= dp^2 + d(d-1)q^2 - 1, \tag{S.42}\\ 
    A_1^B &= A_1^A + 2d(d-1)pq\cos^2(\theta) + 2d\binom{d-1}{2}q^2\cos^4(\theta). \tag{S.43}
\end{align}
We notice that for $p\in [1/d,1]$, $A_1^A$ varies between $0$ and $d-1$.
In fact, for fixed $p$ and $\theta \in [0, \pi/2]$, $A_1^B$ ranges from
$d-1$ to $A_1^A$.
This covers the whole region of allowed values with $A_1^A \leq A_1^B$, as displayed in Fig.~\ref{fig:bipartite_family}. For the other half, one can swap the parties of $\varrho(p,\theta)$.
\end{proof}

\begin{remark}
The geometrical expression of Eq.~(\ref{obs3-twoquditscriterion}) is displayed by
Figs.\,\ref{fig:bipartite_family} and \ref{twoquditsthreedim}.
Also, for $\theta = {\pi}/2$, i.e., $A_1^A = A_1^B$,
one obtains the relation
\begin{align}
A_2 \leq d-1 + \frac{d-2}{2} A_1, \tag{S.44}
    \label{obs3-twoqudits}
\end{align}
which is also expressed geometrically in Fig.\,\ref{twoqudits}.
In the next subsection, we will discuss how to construct the polytope 
of all admissible values of $A_1^A, A_1^B$, and $A_2$, as shown in Figs.\,\ref{twoquditsthreedim} and \ref{twoqudits}.
\end{remark}

\subsection*{A7. Characterization of two-qudit states and discussion of 
the separability criterion}
First, let us characterize two-qudit states using
sector lengths.
This characterization can be useful for understanding the two-qudit
separability criterion geometrically.
The previous work \cite{Nikolai2020} has illustrated the set of 
admissible $(A_1, A_2)$ pairs in two-qubit systems, we will 
generalize it to two-qudit systems.
We begin by recalling that any two-qudit state can be written as
\be \label{blochreptwoqditst}
\varrho_{AB} = \frac{1}{d^2} \sum_{i,j=0}^{d^2-1} t_{ij} \lambda_i \otimes \lambda_j
= \frac{1}{d^2} \left( \eins^{\otimes 2} + P_1 + P_2 \right), \tag{S.45}
\ee
where the Hermitian operators $P_k$
for $k=1,2$ denote the sum of all terms
in the basis element weight $k$.
The relation (\ref{eq-sector-purity}) allows us to translate the purity bound $\mathrm{tr}(\varrho_{AB}^2)\leq 1$ to
\begin{align} \label{ob7-puritybound}
   1+ A_1 + A_2 \leq d^2. \tag{S.46}
\end{align}
Remember that $A_1 = A_1^A + A_1^B$.
As examples of pure states, consider product states $\ket{{\mathrm{prod}}_j}=\ket{jj}$
with the computational basis for $j=0,1,\ldots,(d-1)$.
The pure product states have $A_1^A = A_1^B = d-1$, $A_1 = 2(d-1)$, and $A_2 =(d-1)^2$.
Also, the maximally entangled state $\ket{\Phi^+_d} = (1/{\sqrt{d}})\sum_{i=0}^{d-1} \ket{ii}$
has $A_1^A = A_1^B = A_1 = 0$ and $A_2 =d^2-1$.
It is important to note that the pure product states and the maximally
entangled state can, respectively, maximize the admissible values of $A_1$ and $A_2$
for all two-qudit states (see Ref.~\cite{Eltschka2020}).
That is, both values of sector lengths give tight upper bounds for all two-qudit states:
$A_1 \leq 2(d-1)$ and $A_2 \leq d^2-1$.
Due to the purity condition (\ref{ob7-puritybound}), any pure two-qudit state must satisfy $A_1 \in [0,2(d-1)]$ and $A_2=d^2-1-A_1$.

To see another constraint on sector lengths, let us introduce
the state inversion \cite{Rungta2001, Eltschka2018} expressed as
\begin{align}
    \tilde{\varrho}_{AB} = \frac{1}{d^2} \left\{(d-1)^2 \eins^{\otimes 2} - (d-1)P_1+ P_2 \right\}. \tag{S.47}
\end{align}
Since $\tilde{\varrho}_{AB}$ is positive,
we have 
\begin{align} \label{stateonvtracefrm}
    0 \leq \mathrm{tr}(\varrho_{AB}\tilde{\varrho}_{AB}). \tag{S.48}
\end{align}
From the relation (\ref{eq-sector-purity})
and the expression (\ref{blochreptwoqditst}), the
condition (\ref{stateonvtracefrm})
leads to the state inversion bound:
\begin{align} \label{ob7-stinbound}
    0 \leq (d-1)^2  - (d-1) A_1 + A_2. \tag{S.49}
\end{align}
Here, if $A_2=0$, then $A_1 \leq d-1$, where equality holds if a state is given by,
for example, $\ket{0}\!\bra{0} \otimes {\eins}/{d}$.

\begin{figure}
\includegraphics[width=0.3\linewidth]{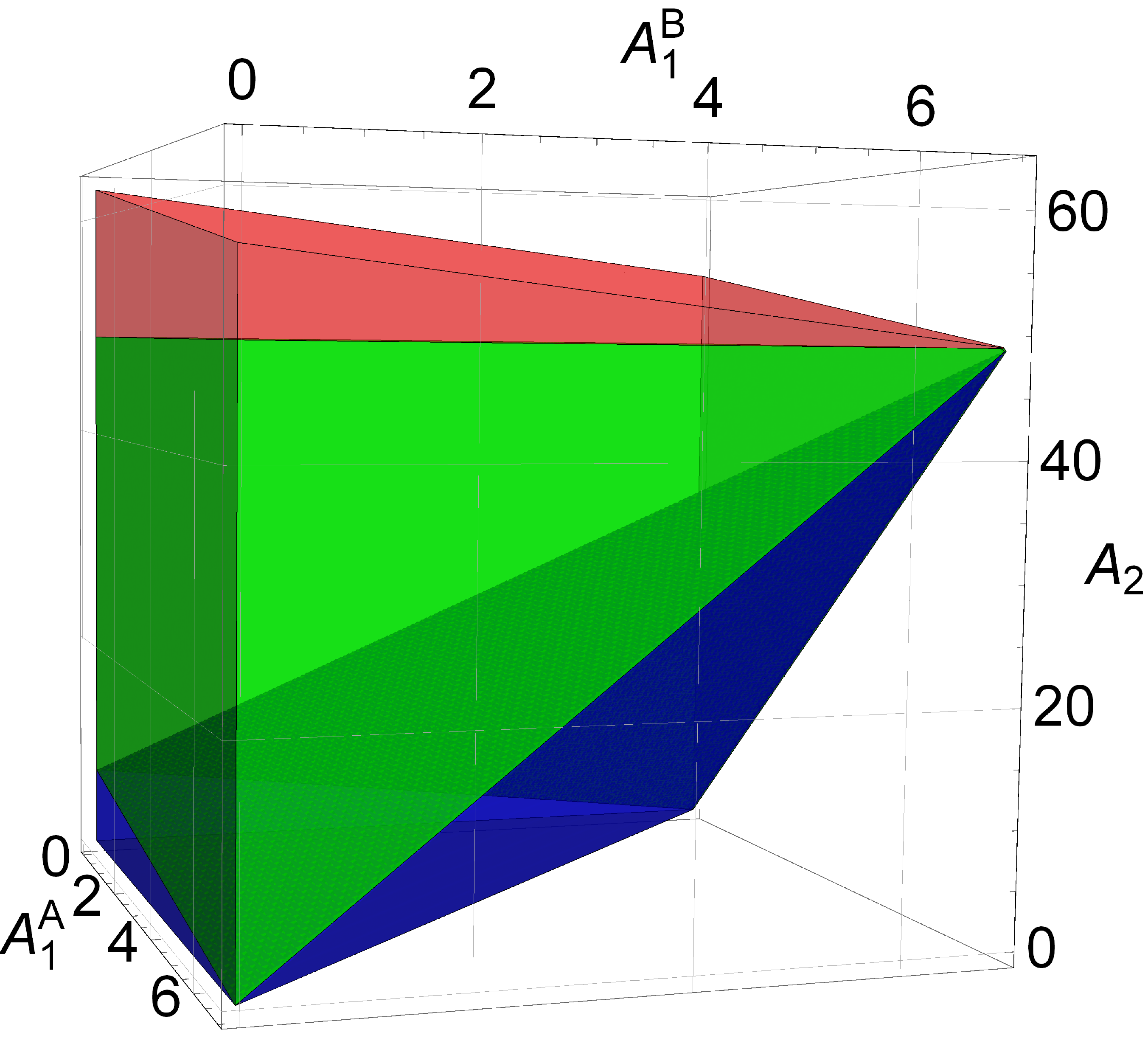}\ \ \ \ \ \ \ \ \ \ \ \ \ \ \ \ \includegraphics[width=0.3\linewidth]{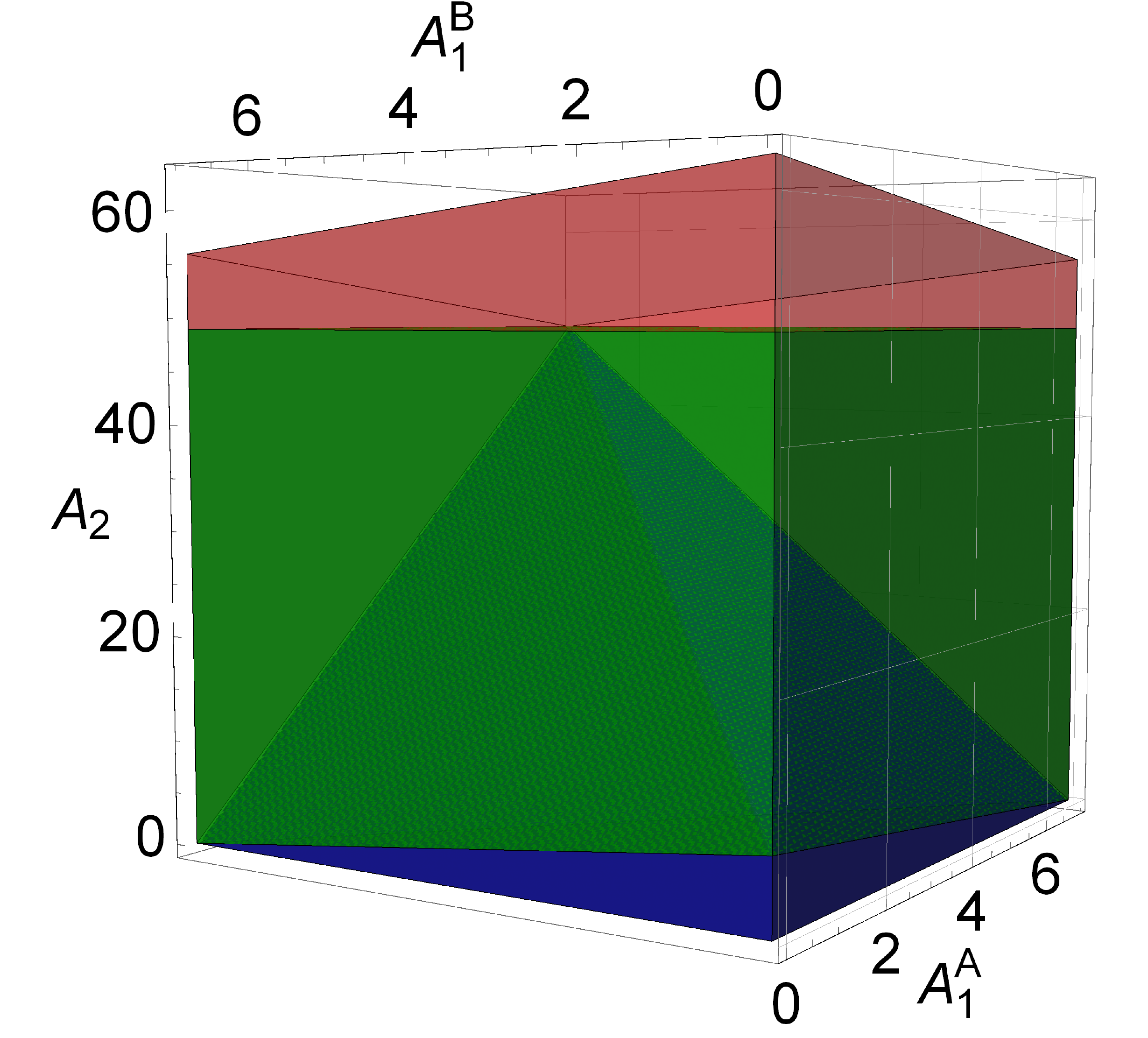}
\caption{Geometry of the state space of $d \otimes d$-dimensional systems
in terms of the quantities $A_1^A, A_1^B,$ and $A_2$, where $d=8$.
The total polytope is the set of all states, characterized by the inequalities
$0 \leq A_1^A, A_1^B \leq d-1$, $0 \leq A_2 \leq d^2-1$,
$A_1^A + A_1^B + A_2 \leq d^2-1$, and $0 \leq (d-1)^2 - (d-1) (A_1^A+A_1^B)+ A_2$.
The separable states are contained in the blue polytope, obeying the
additional constraint in Eq.~(\ref{obs3-twoquditscriterion}).
The red area corresponds to the states violating a previously known criterion
for separable states, $A_2 \leq (d-1)^2$ \cite{Tran2016}.
Thus, the green area marks the improvement coming from Observation 3
compared with the previous result.}
\label{twoquditsthreedim}
\end{figure}

\begin{figure}
    \centering
    \includegraphics[width=0.35\columnwidth]{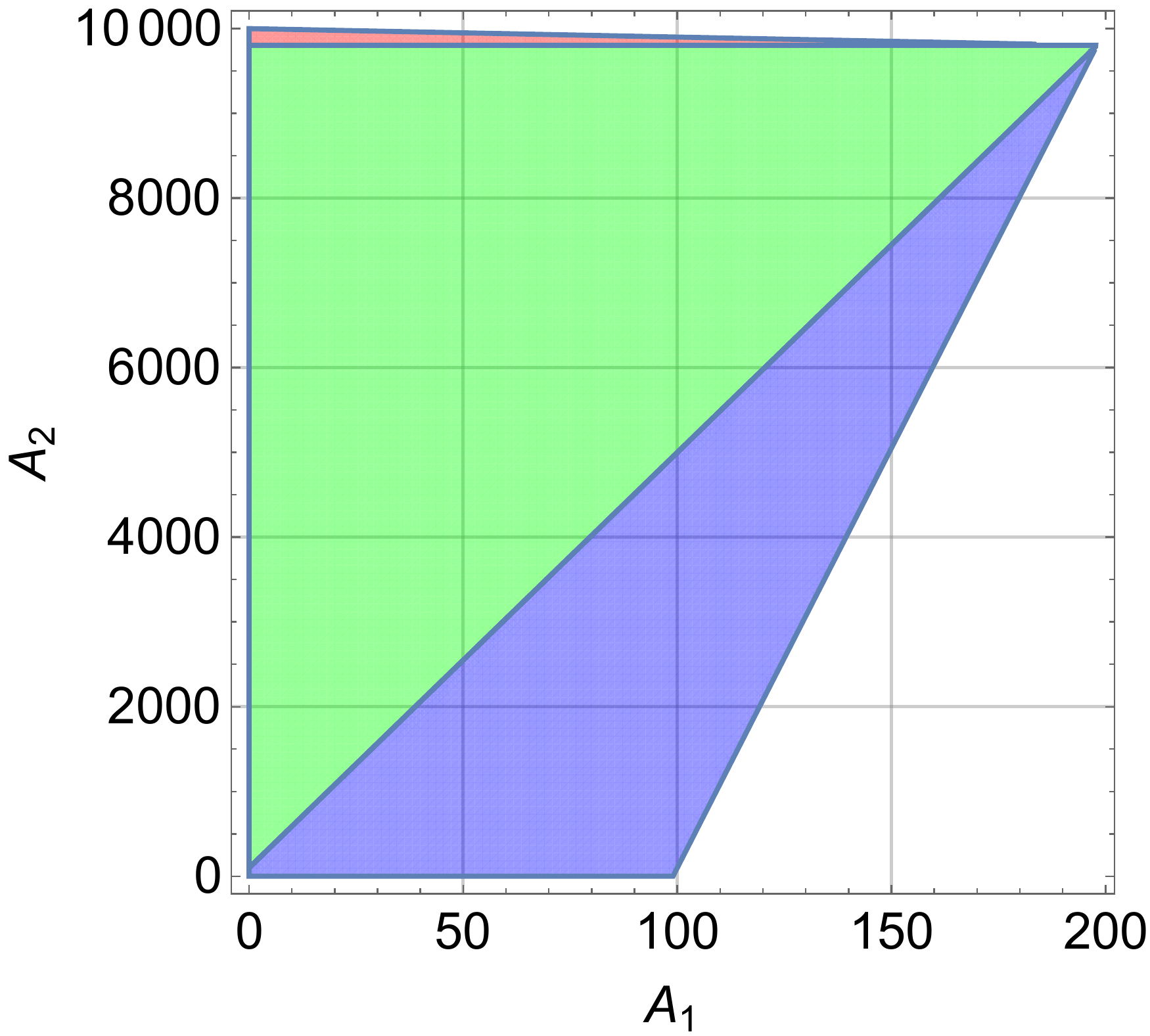}
    \caption{
Geometry of the state space of $d \otimes d$-dimensional systems
in terms of the second moments $A_1$ and $A_2$, where $d=100$.
The total figure is the set of all states, characterized by
the same inequalities with Fig.~\ref{twoquditsthreedim}
in the case $A_1^A = A_1^B$.
The separable states are contained in the blue area,
obeying the additional bound (\ref{obs3-twoqudits}).
The red are corresponds to the state violating a previously known criterion
for separable states, $A_2 \leq (d-1)^2$ \cite{Tran2016}.
Thus, the green area marks the improvement of the criterion in Observation 3
compared with the previous result.
}
\label{twoqudits}
\end{figure}

In conclusion, we obtained the tight four bounds:
$A_1 \leq 2(d-1)$, $A_2 \leq d^2-1$, and Eqs.~(\ref{ob7-puritybound}, \ref{ob7-stinbound}).
These linear constraints on $A_1$ and $A_2$ allow us to ensure the positivity of two-qudit states and to find the total set of their admissible values.
The geometrical expressions are displayed in
Figs.\,\ref{twoquditsthreedim} and \ref{twoqudits}.

Next, let us consider entanglement detection of
two-qudit states.
One existing criterion states that any
two-qudit separable state obeys
$A_2 \leq(d-1)^2$ \cite{Tran2016}, where an
example of states obeying $A_2 = (d-1)^2$
is $\ket{{\mathrm{prod}}_j}$.
In particular,
the maximally entangled state
$\ket{\Phi^+_d}$ maximally violates this inequality.
Now, we look at the gap between the maximally
entangled state and the pure product state
\begin{align}
\frac{A_2(\Phi^+_d)}{A_2({\mathrm{prod}}_j)}
    = \frac{d^2-1}{(d-1)^2} \to 1 \tag{S.50}
\end{align}
for large $d$. 
This scaling tells us that the simple
criterion cannot be useful in very
high-dimensional systems. On the
other hand, the criteria
(\ref{obs3-twoquditscriterion}, \ref{obs3-twoqudits})
allow us to detect entanglement much more powerfully
than the existing criterion,
since they are expressed as the tilted bounds geometrically in
Figs.\,\ref{twoquditsthreedim} and \ref{twoqudits}.
To see that, we consider the two-qudit isotropic state:
\begin{align} \label{isotrpoicstatedbyd}
\varrho_{\mathrm{iso}}=    p\ket{\Phi^+_d}\!\bra{\Phi^+_d} + \frac{1-p}{d^2} \eins^{\otimes 2}, \tag{S.51}
\end{align} 
which has $(A_1, A_2) = (0, p^2(d^2-1))$.
The existing criterion $A_2\leq (d-1)^2$ detects this state as entangled for $p>\sqrt{d-1}/\sqrt{d+1}$, while Eq.~(\ref{obs3-twoqudits})
detects it already for $p > 1/\sqrt{d+1}$, and the state is known to be entangled iff $p > 1/(d+1)$.


\section*{Appendix B: Moments of random correlations in higher-dimensional systems}

In Appendix B, we first evaluate the expressions of $\mathcal{S}_{AB}^{(r)}$ for $r=2$ and $r=4$, originating from the orthogonal average, in terms of the correlation matrix. The calculations are stated rather explicitly here to highlight the methods used. Second, we show that a certain choice of observables yields the same results when performing random unitary measurements instead, i.e., evaluating $\RR_{AB}^{(2)}$ and $\RR_{AB}^{(4)}$, where we skip the detailed calculations, as they  follow similar lines as the detailed orthogonal ones.

\subsection*{B1. Evaluation of $\mathcal{S}_{AB}^{(2)}$ and $\mathcal{S}_{AB}^{(4)}$
for a given state}
In the following, we will show that the moments defined in Eq.~(\ref{eq-newmoments}) 
in the main text are given by
\begin{align} \label{evaluateds2}
    \mathcal{S}^{(2)}_{AB} 
&= V\sum_{i,j} t_{ij}^2, \tag{S.52}\\
   \mathcal{S}^{(4)}_{AB} 
&=  W\left\{
3 \sum_{i,j} t_{ij}^4
+ 3 \sum_{i,j,k, i\neq j} t_{ik}^2 t_{jk}^2
+ 3 \sum_{i,j,k, i\neq j} t_{ki}^2 t_{kj}^2
+ \sum_{i,j,k,l, i\neq k, j\neq l} \left( t_{ij}^2 t_{kl}^2
+ 2 t_{ij}t_{il}t_{kj}t_{kl}\right)
\right\}, \tag{S.53} \label{evaluateds4}
\end{align}
where the $t_{ij}$ are the coefficients from Eq.~(\ref{eq-twoqudits}) in the main text and
\begin{align}
    V=\frac{1}{(d-1)^2}, \ \ \ \ 
    W = \frac{1}{3\left(d-1\right)^4}. \tag{S.54}
\end{align}

Let us substitute the two-qudit state (\ref{eq-twoqudits}) 
in the main text into the moments
$\mathcal{S}^{(r)}_{AB}$. In the following, we always 
add a normalization $N=N(r,d)$, which is later chosen
such that $\mathcal{S}^{(r)}_{AB}=1$ for pure product 
states, see also Eq.~(\ref{eq-nrd-norm}) below. 
Then we have 
\begin{align} \nonumber
\mathcal{S}^{(r)}_{AB} 
&= N\int d\bm{\alpha}_1 \int d\bm{\alpha}_2 \,
    \left\{
    \mathrm{tr}
    \left[
    \left(\frac{1}{d^2} \sum_{i,j=0}^{d^2-1} t_{ij}
    \lambda_i \otimes \lambda_j
    \right)
    \bm{\alpha}_1 \cdot \bm{\lambda} \otimes \bm{\alpha}_2 \cdot \bm{\lambda}
    \right]
    \right\}^r\\ \nonumber
&= \frac{N}{d^{2r}}
   \int d\bm{\alpha}_1 \int d\bm{\alpha}_2 \,
   \left\{
   \sum_{i,j=1}^{d^2-1} t_{ij}
   \mathrm{tr} \left[\lambda_i \bm{\alpha}_1 \cdot \bm{\lambda} \right] \cdot
   \mathrm{tr} \left[\lambda_j \bm{\alpha}_2 \cdot \bm{\lambda} \right]
   \right\}^r \\ \nonumber
&= N\int d\bm{\alpha}_1 \int d\bm{\alpha}_2 \,
   \left[
   \sum_{i,j=1}^{d^2-1} t_{ij} {\alpha}_1^{(i)} {\alpha}_2^{(j)}
   \right]^r \\ \nonumber
&= N\int d\bm{\alpha}_1 \int d\bm{\alpha}_2 \,
   \sum_{r_{i,j}} \frac{r!}{r_{1,1}!\cdots r_{d^2-1,\,d^2-1}!}
   \prod_{i,j=1}^{d^2-1}
   \left[
   t_{ij} {\alpha}_1^{(i)} {\alpha}_2^{(j)}
   \right]^{r_{i,j}} \\ \nonumber
&= N \sum_{r_{i,j}} \frac{r!}{r_{1,1}!\cdots r_{d^2-1,\,d^2-1}!}
   \prod_{i,j=1}^{d^2-1} t_{ij} ^{r_{i,j}}
   \int d\bm{\alpha}_1 \, \prod_{i=1}^{d^2-1} \left[{\alpha}_1^{(i)} \right]^{\sum_{j=1}^{d^2-1} r_{i,j} }
   \int d\bm{\alpha}_2 \, \prod_{j=1}^{d^2-1} \left[{\alpha}_2^{(j)} \right]^{\sum_{i=1}^{d^2-1} r_{i,j} } \\ \nonumber
&= N \sum_{r_{i,j}} \frac{r!}{r_{1,1}!\cdots r_{d^2-1,\,d^2-1}!}
   \prod_{i,j=1}^{d^2-1} t_{ij} ^{r_{i,j}}
   \int d\bm{\alpha}_1 \, \prod_{i=1}^{d^2-1} \left[{\alpha}_1^{(i)} \right]^{a_i}
   \int d\bm{\alpha}_2 \, \prod_{j=1}^{d^2-1} \left[{\alpha}_2^{(j)} \right]^{a^{\prime}_j } \\
   \label{stndardexpsiionrr}
&= 4N \sum_{r_{i,j}} \frac{r!}{r_{1,1}!\cdots r_{d^2-1,\,d^2-1}!}
   \prod_{i,j=1}^{d^2-1} t_{ij} ^{r_{i,j}}
   B(b_1,b_2,\ldots,b_{d^2-1}) B(b^{\prime}_1,b^{\prime}_2,\ldots,b^{\prime}_{d^2-1}),  \tag{S.55}
\end{align}
where we use
$\mathrm{tr} (\bm{\alpha}_i \cdot \bm{\lambda}) =0$
and the multinomial theorem
\begin{align}
    \left(\sum_{i=1}^n x_i\right)^r = \sum_{r_1+r_2+ \cdots +r_n =r} \frac{r!}{r_1!r_2!\cdots r_n!} \prod_{i=1}^{n}x_i^{r_i}. \tag{S.56}
\end{align}
The sum $\sum_{r_{i,j}}$ means
$\sum_{r_{1,1} + \cdots + r_{1,\, d^2-1} + r_{2,1} + \cdots + r_{d^2-1,\,d^2-1} = r}$.
We define that
$a_i = \sum_{j=1}^{d^2-1} r_{i,j}$,
$a^{\prime}_j = \sum_{i=1}^{d^2-1} r_{i,j}$,
$b_i = (a_i +1)/2$, and
$b^{\prime}_j = (a^{\prime}_j +1)/2$.
Note that in general, the integral over the
$n$-dimensional unit sphere is written as
$2B(\beta_1, \beta_2, \ldots, \beta_n)$
(see Ref.~\cite{Folland2001}),
where $B(\beta_1, \beta_2, \ldots, \beta_n)$
denotes the multi-variable beta function, for
$\beta_i = (\alpha_i +1)/2$ and the gamma function
$\Gamma(\beta_i)$, given by
\begin{align}\label{eq:Bfunction}
    B(\beta_1, \beta_2, \ldots, \beta_n)
    = \frac{\Gamma(\beta_1) \Gamma(\beta_2) \cdots \Gamma(\beta_{n})}
    {\Gamma(\beta_1 + \beta_2 + \ldots + \beta_{n})}. \tag{S.57}
\end{align}
This integral vanishes if any of $\alpha_i$
is odd.

\subsubsection*{Case of $r=2$}
Let us evaluate the moment at $r=2$.
The condition $r=2$ means
$r_{1,1} + \cdots + r_{d^2-1,\,d^2-1} = 2$.
To make this more explicit,
we introduce the square matrix $\bm{R}$
\begin{align}
\bm{R} = 
\begin{pmatrix}
r_{1,1}&r_{1,2}&\cdots&r_{1,\, d^2-1} \\ 
r_{2,1}&r_{2,2}&\cdots&r_{2,\, d^2-1}\\
\vdots&\vdots  &\ddots&\vdots\\
r_{d^2-1,1}&r_{d^2-1,2}&\cdots&r_{d^2-1,\,d^2-1}
\end{pmatrix}. \tag{S.58}
\end{align}
Recall here that
$a_i = \sum_{j=1}^{d^2-1} r_{i,j}$ and
$a^{\prime}_j = \sum_{i=1}^{d^2-1} r_{i,j}$.
Thus, $a_i$ and $a^{\prime}_j$ respectively
correspond to the $i$-th row vector
and the $j$-th column vector of the matrix
$\bm{R}$.

There are two candidates that satisfy the
condition
$r_{1,1} + \cdots + r_{d^2-1,\,d^2-1} = 2$:
\begin{itemize}
    \item[(1)]
    one of elements is equal to $2$ and all
    other elements are zero, that is,
    fixed $r_{\alpha, \beta} = 2$ and
    $r_{k,l \neq \alpha, \beta} =0$.
    An example is given by
\begin{align} \label{requal2case}
\bm{R} = 
\begin{pmatrix}
2&0&\cdots&0 \\ 
0&0&\cdots&0\\
\vdots&\vdots  &\ddots&\vdots\\
0&0&\cdots&0
\end{pmatrix}. \tag{S.59}
\end{align}

    \item[(2)]
    two of elements are equal to $1$ and all
    other elements are zero, that is,
    fixed
    $r_{\alpha, \beta} = r_{\gamma, \delta} =1$
    and all other $r_{k,l} =0$.
    Examples are 
\begin{align} 
\bm{R} = 
\begin{pmatrix}
0&1&\cdots&0 \\ 
0&1&\cdots&0\\
\vdots&\vdots  &\ddots&\vdots\\
0&0&\cdots&0
\end{pmatrix},
\ \
\begin{pmatrix}
1&1&\cdots&0 \\ 
0&0&\cdots&0\\
\vdots&\vdots  &\ddots&\vdots\\
0&0&\cdots&0
\end{pmatrix},
\ \
\begin{pmatrix}
1&0&\cdots&0 \\ 
0&1&\cdots&0\\
\vdots&\vdots  &\ddots&\vdots\\
0&0&\cdots&0
\end{pmatrix}. \tag{S.60}
\end{align}
\end{itemize}

In any case of the candidate (2), either or
both of $a_i$ and $a^{\prime}_j$
are always $1$, that is, odd.
This results in the vanishing
of the integral over the sphere.
Accordingly, it is sufficient to
focus only on the candidate (1) in Eq.~(\ref{requal2case}).
Concerning the expression of moments
(\ref{stndardexpsiionrr}), we have
\begin{align} \nonumber
    &\frac{2!}{0! \cdots 2! \cdots 0!} = 1, \ \
    \prod_{i,j=1}^{d^2-1} t_{ij} ^{r_{i,j}}=t_{\alpha \beta}^2,\\ \nonumber
    &a_{\alpha} = \sum_j r_{\alpha,j} = 2, \ a_{i \neq \alpha} = 0, \ \
    a^{\prime}_{\beta} = \sum_i r_{i,\beta} = 2, \ a^{\prime}_{i \neq \beta} =0,\\ \nonumber
    &b_{\alpha} =\frac{3}{2}, \ b_{i \neq \alpha} = \frac{1}{2}, \ \
    b^{\prime}_{\beta} =\frac{3}{2}, \ b^{\prime}_{i \neq \beta}=\frac{1}{2},\\ 
    &B(b_1, \ldots, b_{\alpha}, \ldots, b_{d^2-1})=B(b^{\prime}_1,\ldots, b^{\prime}_{\beta}, \ldots,b^{\prime}_{d^2-1})
    =B\left(\frac{1}{2},\ldots, \frac{3}{2}, \ldots, \frac{1}{2} \right)
        =\frac{(\sqrt{\pi})^{d^2-1}}{2\Gamma \left( \frac{d^2+1}{2}\right)}. \tag{S.61}
\end{align}
Then we have
\begin{align} 
    \mathcal{S}^{(2)}_{AB} 
= V\sum_{i,j} t_{ij}^2, \tag{S.62}
\end{align}
where
\begin{align} 
    V=4N \frac{\pi^{d^2-1}}{4 \left[\Gamma \left( \frac{d^2+1}{2}\right)\right]^2} = \frac{1}{(d-1)^2}. \tag{S.63}
\end{align}

\subsubsection*{Case of $r=4$}
Let us evaluate the moment at $r=4$.
The condition $r=4$ means
$r_{1,1} + \cdots + r_{d^2-1,\,d^2-1} = 4$.
There are several candidates that satisfy
the condition
$r_{1,1} + \cdots + r_{d^2-1,\,d^2-1} = 4$.
In the following, we will only describe the three
candidates with nonzero values of the
integral over the sphere.
\begin{itemize}
    \item[(1)]
    one of the elements is equal to $4$ and all
    other elements are zero, that is,
    fixed $r_{\alpha, \beta} = 4$ and
    $r_{k,l \neq \alpha, \beta} =0$.
    An example is
\begin{align} 
\bm{R} = 
\begin{pmatrix}
4&0&\cdots&0 \\ 
0&0&\cdots&0\\
\vdots&\vdots  &\ddots&\vdots\\
0&0&\cdots&0
\end{pmatrix}. \tag{S.64}
\end{align}

    \item[(2)]
    two of elements are equal to $2$ and
    all other elements are zero, that is,
    fixed
    $r_{\alpha, \beta} = r_{\gamma, \delta} =2$
    and others $r_{k,l} =0$.
    Examples are divided into three types.
\begin{align} \label{abccases}
\bm{R} = 
\begin{pmatrix}
0&2&\cdots&0 \\ 
0&2&\cdots&0\\
\vdots&\vdots  &\ddots&\vdots\\
0&0&\cdots&0
\end{pmatrix},
\ \
\begin{pmatrix}
2&2&\cdots&0 \\ 
0&0&\cdots&0\\
\vdots&\vdots  &\ddots&\vdots\\
0&0&\cdots&0
\end{pmatrix},
\ \
\begin{pmatrix}
2&0&\cdots&0 \\ 
0&2&\cdots&0\\
\vdots&\vdots  &\ddots&\vdots\\
0&0&\cdots&0
\end{pmatrix}. \tag{S.65}
\end{align}
We call these (a), (b), and (c) cases, respectively.

    \item[(3)]
    four of elements are equal to $1$
    and all other elements are zero,
    that is, fixed
    $r_{\alpha, \beta} = r_{\gamma, \delta} = r_{\epsilon, \zeta} = r_{\eta, \theta} = 1$
    and others are zero.
    An example is 
\begin{align} 
\bm{R} = 
\begin{pmatrix}
1&1&\cdots&0 \\ 
1&1&\cdots&0\\
\vdots&\vdots  &\ddots&\vdots\\
0&0&\cdots&0
\end{pmatrix}. \tag{S.66}
\end{align}
\end{itemize}

\textit{Candidate (1).}
Let us consider the candidate (1). For fixed $r_{\alpha, \beta} = 4$
and $r_{k,l \neq \alpha, \beta} =0$, we have
\begin{align} \nonumber
    &a_{\alpha} = 4, \ a^{\prime}_{\beta} = 4,\ \
    b_{\alpha} =\frac{5}{2}, \ b^{\prime}_{\beta} =\frac{5}{2},\\ 
    &B(b_1, \ldots, b_{\alpha}, \ldots, b_{d^2-1})=B(b^{\prime}_1,\ldots, b^{\prime}_{\beta}, \ldots,b^{\prime}_{d^2-1})
    =B\left(\frac{1}{2},\ldots, \frac{5}{2}, \ldots, \frac{1}{2} \right)
    =\frac{3(\sqrt{\pi})^{d^2-1}}{4\Gamma \left( \frac{d^2+3}{2}\right)}. \tag{S.67}
\end{align}

Therefore, the corresponding term is given by
\begin{align} 
    4N \frac{9 \pi^{d^2-1}}{16 \left[\Gamma \left( \frac{d^2+3}{2}\right)\right]^2} \sum_{i,j} t_{ij}^4. \tag{S.68}
\end{align}

\textit{Candidate (2).}
Let us consider the candidate (2).
For fixed
$r_{\alpha, \beta} = r_{\gamma, \delta} = 2$,
we have the three types (a), (b), and (c),
as the three examples described in
(\ref{abccases}):
\begin{itemize}
    \item[(a)] $\alpha \neq \gamma$ and $\beta = \delta$.
    \item[(b)] $\alpha = \gamma$ and $\beta \neq \delta$.
    \item[(c)] $\alpha \neq \gamma$ and $\beta \neq \delta$.
\end{itemize}

For the type (a), we have
\begin{align} \nonumber
    &a_{n} = 2, \ \mathrm{for} \, n=\alpha, \gamma, \ a^{\prime}_{\beta} = 4,\ \
    b_{n} =\frac{3}{2}, \ b^{\prime}_{\beta} =\frac{5}{2},\\ \nonumber
    &B(b_1, \ldots, b_{\alpha}, \ldots, b_{\gamma}, \ldots, b_{d^2-1})
    =B\left(\frac{1}{2},\ldots, \frac{3}{2}, \ldots, \frac{3}{2}, \ldots, \frac{1}{2} \right)
    =\frac{(\sqrt{\pi})^{d^2-1}}{4\Gamma \left( \frac{d^2+3}{2}\right)},\\ 
    &B(b^{\prime}_1,\ldots, b^{\prime}_{\beta}, \ldots,b^{\prime}_{d^2-1})
    =B\left(\frac{1}{2},\ldots, \frac{5}{2}, \ldots, \frac{1}{2} \right)
    =\frac{3(\sqrt{\pi})^{d^2-1}}{4\Gamma \left( \frac{d^2+3}{2}\right)}. \tag{S.69}
\end{align}
Moreover, to avoid the over-counting of summation, we multiply it by $1/2$ in order to be able to write the contribution as the following sum:
\begin{align} 
    4N \times 6 \times \frac{1}{2} \times \frac{3 \pi^{d^2-1}}{16 \left[\Gamma \left( \frac{d^2+3}{2}\right)\right]^2} \sum_{i,j,k, i\neq j} t_{ik}^2 t_{jk}^2. \tag{S.70}
\end{align}

For the type (b), we have the similar result with (a):
\begin{align} 
    4N \times 6 \times \frac{1}{2} \times \frac{3 \pi^{d^2-1}}{16 \left[\Gamma \left( \frac{d^2+3}{2}\right)\right]^2} \sum_{i,j,k, i\neq j} t_{ki}^2 t_{kj}^2. \tag{S.71}
\end{align}

For the type (c), we have
\begin{align} \nonumber
    &a_{n} = 2, \ a^{\prime}_{m} = 2, \ \mathrm{for} \, n=\alpha, \gamma, \ m=\beta, \delta, \ \
    b_{n} =\frac{3}{2}, \ b^{\prime}_{m} =\frac{3}{2},\\ 
    &B(b_1, \ldots, b_{\alpha}, \ldots, b_{\gamma}, \ldots, b_{d^2-1}) =B(b^{\prime}_1,\ldots, b^{\prime}_{\beta}, \ldots, b^{\prime}_{\delta}, \ldots, b^{\prime}_{d^2-1})
    =B\left(\frac{1}{2},\ldots, \frac{3}{2}, \ldots, \frac{3}{2}, \ldots, \frac{1}{2} \right)
    =\frac{(\sqrt{\pi})^{d^2-1}}{4\Gamma \left( \frac{d^2+3}{2}\right)}. \tag{S.72}
\end{align}
Therefore, the corresponding term is given by
\begin{align} 
    4N \times 6 \times \frac{1}{2} \times \frac{\pi^{d^2-1}}{16 \left[\Gamma \left( \frac{d^2+3}{2}\right)\right]^2} \sum_{i,j,k,l, i\neq k, j\neq l} t_{ij}^2 t_{kl}^2. \tag{S.73}
\end{align}

\textit{Candidate (3).}
Let us consider the candidate (3). For fixed
$r_{\alpha, \beta} = r_{\gamma, \delta} = r_{\epsilon, \zeta} = r_{\eta, \theta} = 1$,
only one case yields finite values of the integral:
$\alpha=\gamma$, $\beta=\zeta$, $\epsilon=\eta$, and $\delta=\theta$.
Here, rewriting the condition as
$r_{\alpha, \beta} = r_{\alpha, \delta} = r_{\epsilon, \beta} = r_{\epsilon, \delta} = 1$,
we have 
\begin{align} \nonumber
    &a_{n} = 2, \ a^{\prime}_{m} = 2, \ \mathrm{for} \, n=\alpha, \epsilon, \ m=\beta, \delta, \ \
    b_{n} =\frac{3}{2}, \ b^{\prime}_{m} =\frac{3}{2},\\    &B(b_1, \ldots, b_{\alpha}, \ldots, b_{\epsilon}, \ldots, b_{d^2-1})
    =B(b^{\prime}_1,\ldots, b^{\prime}_{\beta}, \ldots, b^{\prime}_{\delta}, \ldots, b^{\prime}_{d^2-1})
    =B\left(\frac{1}{2},\ldots, \frac{3}{2}, \ldots, \frac{3}{2}, \ldots, \frac{1}{2} \right)
    =\frac{(\sqrt{\pi})^{d^2-1}}{4\Gamma \left( \frac{d^2+3}{2}\right)}. \tag{S.74}
\end{align}
Moreover, to avoid the over-counting of summation, we multiply it by $1/4$.
Therefore, the corresponding term is given by
\begin{align} 
    4N \times 24 \times \frac{1}{4} \times \frac{\pi^{d^2-1}}{16 \left[\Gamma \left( \frac{d^2+3}{2}\right)\right]^2} \sum_{i,j,k,l, i\neq k, j\neq l} t_{ij}t_{il}t_{kj}t_{kl}. \tag{S.75}
\end{align}

According to the candidates (1), (2), and  (3), we
finally arrive at
\begin{align} 
   \mathcal{S}^{(4)}_{AB}  = W\left\{
3 \sum_{i,j} t_{ij}^4
+ 3 \sum_{i,j,k, i\neq j} t_{ik}^2 t_{jk}^2
+ 3 \sum_{i,j,k, i\neq j} t_{ki}^2 t_{kj}^2
+ \sum_{i,j,k,l, i\neq k, j\neq l} \left( t_{ij}^2 t_{kl}^2
+ 2 t_{ij}t_{il}t_{kj}t_{kl}\right)
\right\}, \tag{S.76}
\end{align}
where
\begin{align} 
    W = 4N \frac{3 \pi^{d^2-1}}{16 \left[\Gamma \left( \frac{d^2+3}{2}\right)\right]^2}
    = \frac{1}{3\left(d-1\right)^4}. \tag{S.77}
\end{align}

\subsection*{B2. Suitable observables in higher dimensions}
Let us discuss the relation between the moments
$\mathcal{R}^{(r)}_{AB}$
and $\mathcal{S}^{(r)}_{AB}$.
In order to explain the difficulties for higher dimensions,
let us focus on qubits first.
Suppose that Alice and Bob locally
perform the measurements $M_A$ and $M_B$ in
random bases parametrized by the unitary
transformations $U_A, U_B \in \mathcal{U}(d)$,
such that
\begin{align}
    &\left\{\ket{u_0}_A =U_A \ket{0}_A, \ket{u_1}_A=U_A \ket{1}_A, \ldots,  \ket{u_{d-1}}_A=U_A \ket{d-1}_A \right\}, \tag{S.78}\\
    &\left\{\ket{u_0}_B =U_B \ket{0}_B, \ket{u_1}_B=U_B \ket{1}_B, \ldots,  \ket{u_{d-1}}_B=U_B \ket{d-1}_B \right\}. \tag{S.79}
\end{align}
In the case of qubits ($d=2$), Alice's (Bob's)
measurement direction corresponds
to a random three-dimensional unit vector
$\bm{u}_A$ ($\bm{u}_B$) chosen uniformly
on the Bloch sphere $\mathcal{S}^2$. Then, the
expectation value is given by
$\mathrm{tr}\left[\varrho_{AB} \sigma_{\bm{u}_A} \otimes \sigma_{\bm{u}_B} \right]$,
where
$\sigma_{\bm{u}} = \bm{u} \cdot \bm{\sigma}$
is the rotated Pauli matrix
with the vector of the usual Pauli matrices
${\bm{\sigma}}=(\sigma_x, \sigma_y, \sigma_z)^{\top}$.
Without loss of generality, one can take the Pauli-$Z$ matrix
$\sigma_z$ as the observables $M_A$ and $M_B$.
Then we can characterize the obtained
distribution via
its moments $\RR^{(r)}_{AB}$
\begin{align} 
    \RR^{(r)}_{AB}
= \int dU_A \int dU_B \left\{
\mathrm{tr}[\varrho_{AB} (U_A \sigma_z U_A^\dagger)
\otimes (U_B \sigma_z U_B^\dagger)]
\right\}^r 
= \frac{1}{(4\pi)^2} \int_{\mathcal{S}^2} d\bm{u}_A \int_{\mathcal{S}^2} d\bm{u}_B
\left[
\mathrm{tr}\left(\varrho_{AB} \sigma_{\bm{u}_A} \otimes \sigma_{\bm{u}_B}
\right) \right]^{r}, \tag{S.80}
\end{align}
where the unitaries are typically chosen
according to the Haar distribution.
For all odd $r$, the moments $\mathcal{R}^{(r)}_{AB}$ vanish,
so the quantities of interest are the moments of even $r$.
Indeed, the second moment $\mathcal{R}^{(2)}_{AB}$ can be
evaluated by a unitary two-design \cite{Dankert2005}:
\begin{align} \label{eq:R2abmoment}
    \mathcal{R}^{(2)}_{AB} = \frac{1}{9}
    \sum_{\bm{e}_A, \bm{e}_B = \bm{e}_1, \bm{e}_2, \bm{e}_3}
    \mathrm{tr}\left[\varrho_{AB} \sigma_{\bm{e}_A} \otimes \sigma_{\bm{e}_B} \right] ^2 
    =\frac{1}{9}\sum_{i,j=1}^{3} t_{i j}^2, \tag{S.81}
\end{align}
where $\{\pm \bm{e}_k \mid k=1,2,3 \}$ are the
orthogonal local directions and the $t_{ij}$ are
two-body correlation coeffients with $1\leq i,j \leq 3$ of $\varrho_{AB}$,
where we call this submatix $T_s$. 
It is important that the moments $\mathcal{R}^{(r)}_{AB}$
are by definition invariant under local unitary transformations $U_A \otimes U_B$.
This property allows us to find a local unitary such that
the matrix $T_s$ can be diagonalized by a orthogonal transformation,
due to the isomorphism between $SO(3)$ and $SU(2)$.

On the other hand, in the case of higher dimensions
($d>2$), there are several problems.
First, the notion of a Bloch sphere is not available.
Due to this fact, not all possible observables are equivalent under randomized unitaries.
Second, for a odd $r$, the moments $\mathcal{R}^{(r)}_{AB}$ 
in Eq.~(\ref{eq-moments1}) in the main text do not vanish,
see Ref.~\cite{Krebsbach2019}.
Third, for $r=2$, the second moments are independent
of the choice of observables
as long as the observables
are traceless (see Theorem 9 in Ref.~\cite{Tran2016}),
while higher moments depend on the choice.
To approach these problems, we make use of the quantities from the previous section, i.e.,
\begin{align} 
   \mathcal{S}^{(r)}_{AB} &=N(r,d)\int_{\mathcal{V}^{d^2-2}} d\bm{\alpha}_1
   \int_{\mathcal{V}^{d^2-2}} d\bm{\alpha}_2
     [\mathrm{tr}
    (\varrho_{AB} \bm{\alpha}_1 \cdot \bm{\lambda} \otimes \bm{\alpha}_2 \cdot \bm{\lambda} )
    ]^r \tag{S.82}\\
   &\propto \int dO_A
   \int dO_B
     \left\{\mathrm{tr}
     \left[
    \sum_{i,j}
    \left((O_A T_s O_B)_{ij} (\lambda_i \otimes \lambda_j)
    \right)
    (\lambda_1 \otimes \lambda_1) \right]
    \right\}^r, \tag{S.83}
\end{align}
where
$O_A, O_B \in SO(d^2-1)$ are non-physical orthogonal matrices and 
the elements of the correlation matrix $T_s$ are given by $(T_s)_{ij} = \mathrm{tr}(\varrho_{AB} \lambda_i \otimes \lambda_j)$.
In addition, the $\bm{\alpha}_i$ denote the
$(d^2-1)$-dimensional unit real vectors
uniformly chosen from on the pseudo Bloch sphere
$\mathcal{V}^{d^2-2}$, and
$\bm{\lambda} = (\lambda_1, \lambda_2, \ldots, \lambda_{d^2-1} )$ is the
vector of Gell-Mann matrices.
Here, $N(r,d)$ is a normalization factor such that
${\mathcal{S}}^{(r)}_{AB}=1$
at pure product states:
\begin{align}
    N(r,d)=\frac {[(d^2+r-3)!!]^2} {(d-1)^r[(r-1)!!]^2[(d^2-3)!!]^2}
    \left(
    \frac{\Gamma(\frac{d^2-1}{2})}{2\sqrt{\pi}^{d^2-1}}
    \right)^2,
    \label{eq-nrd-norm} \tag{S.84}
\end{align}
where, for a positive number $n$, $n!!$ is the double
factorial and $\Gamma (n)$ is the gamma function.
The moments $\mathcal{S}^{(r)}_{AB}$ are analytically calculable,
so we take $\mathcal{S}^{(r)}_{AB}$ as the starting point for our discussion.

The question here is whether
it is possible to find observables such that
$\mathcal{R}^{(r)}_{AB}$ coincides with
$\mathcal{S}^{(r)}_{AB}$, up to a constant.
While the observables $M_A$ and $M_B$ do not have to be diagonal,
they can be assumed to be diagonal in the unitary group averaging.
Now let us consider a diagonal observable
such that $M_A=M_B=M_d$.
Then, we are in a position to
present the suitable choice of $M_d$
for the coincidence between $\mathcal{S}^{(r)}_{AB}$
and $\mathcal{R}^{(r)}_{AB}$.

\noindent
{\bf Observation 9.}
{\it
In $d$-dimensional quantum systems where $d$ is odd,
let the diagonal observable $M_d$ be given by
\kommentar{
\begin{align}
    {M_d}=\frac{\sqrt{d}}{
    \mathrm{tr}
    \left(M^2\right) } M,
\end{align}
where
\begin{align}\nonumber
    M&= \eins_{(d-1)/2} \oplus (y) \oplus 0_{(d-1)/2} - \frac{2y+d-1}{2d} \eins_{d}\\
          &= \operatorname{diag}(\underbrace{1,\ldots,1}_{(d-1)/2},y,\underbrace{0,\ldots,0}_{(d-1)/2}) - \frac{2y+d-1}{2d}\eins_{d},
          \\
          y &= \frac12\left[1\pm \sqrt{1 + \frac{d+3+\sqrt{d^3+3d^2+d+3}}{d-2}}\right],
    \label{eq:ysoldodd}
\end{align}
}
\begin{align}
    M_d = \operatorname{diag}(\underbrace{\alpha_+,\ldots,\alpha_+}_{(d-1)/2},\beta_y,\underbrace{\alpha_-,\ldots,\alpha_-}_{(d-1)/2}), \tag{S.85}
\end{align}
where
\begin{align}
\alpha_\pm &= \frac{\pm d-2y+1}{\sqrt{(d-1)[(2y-1)^2 + d)]}}, \tag{S.86}\\
\beta_y &= -\sqrt{ \frac{(d-1)(2y-1)^2}{(2y-1)^2 + d}}, \tag{S.87}\\
y &= \frac12\left[1 - \sqrt{1 + \frac{d+3+\sqrt{d^3 +3d^2 + d + 3}}{d-2}}\right], \tag{S.88}
\label{eq:ysoldodd}
\end{align}
and $\mathrm{tr}(M_d)=0$ and $\mathrm{tr}(M_d^2)=d$.
Then, measuring the observable $M_d$
yields 
\begin{align}
    \mathcal{S}^{(2)}_{AB}
    =(d+1)^2 \mathcal{R}^{(2)}_{AB}, \ \ \ \ \ \ \ \ 
    \mathcal{S}^{(4)}_{AB}
    =\frac{(d+1)^2(d^2+1)^2}{9(d-1)^2} \mathcal{R}^{(4)}_{AB}. \tag{S.89}
\end{align}
}

\begin{proof}
Analogous to the calculation in the case of
$\mathcal{S}^{(r)}_{AB}$ shown in Eq.~(\ref{stndardexpsiionrr}),
after some lengthy calculation and using the fact
that $M_d$ is traceless, we obtain
\begin{align}\label{eq:rexprint}
    \mathcal{R}^{(r)}_{AB}
    = \sum_{r_{i,j}} \frac{r!}{r_{1,1}!\cdots r_{d^2-1,\,d^2-1}!}
   \prod_{i,j=1}^{d^2-1} t_{ij} ^{r_{i,j}}
  \int dU_A \prod_{i=1}^{d^2-1} \mathrm{tr}[U_A M_d U_A^\dagger\lambda_i]^{a_i}
   \int dU_B \prod_{j=1}^{d^2-1} \mathrm{tr}[U_B M_d U_B^\dagger\lambda_j]^{a^{\prime}_j }, \tag{S.90}
\end{align}
where the sum spans over all non-negative integer
assignments to the $r_{i,j}$
such that $\sum_{i,j=1}^{d^2-1} r_{i,j} = r$, and
$a_i = \sum_{j=1}^{d^2-1} r_{i,j}$, $a_j^\prime = \sum_{i=1}^{d^2-1} r_{i,j}$.

We start with the discussion of the case $r=4$, where
we focus on one of the integrals and evaluate it for all possible exponent
vectors $\vec{a} = (a_1,\ldots,a_{d^2-1})$.
As all the entries are positive integers that
sum to $4$, there are five families of vectors to
be considered:
\begin{itemize}
    \item[(a)] $a_k = 4$, $a_{l\neq k} = 0$,
    \item[(b)] $a_k = 3$, $a_l = 1$, $a_{m\neq k,l} = 0$,
    \item[(c)] $a_k = 2$, $a_l = 2$,  $a_{m\neq k,l} = 0$,
    \item[(d)] $a_k = 2$, $a_l = 1$, $a_m = 1$, $a_{n\neq k,l,m} = 0$,
    \item[(e)] $a_k = 1$, $a_l = 1$, $a_m = 1$, $a_n = 1$, $a_{o\neq k,l,m,n} = 0$.
\end{itemize}
We aim to prove that the whole integral coincides
with the one obtained from integration over the
orthogonal group. To that end, we compare the integrals occurring in Eq.~(\ref{eq:rexprint}) with those in Eq.~(\ref{stndardexpsiionrr}). In particular, we try to tweak the observable such that for all vectors $\vec{a}$ with $\sum a_i = 4$,
\begin{align}
    \int dU_A \prod_{i=1}^{d^2-1} \mathrm{tr}[U_A M_d U_A^\dagger\lambda_i]^{a_i} = M B(b_1, b_2,\ldots, b_{d^2-1}), \tag{S.91}
\end{align}
where $B$ is defined in Eq.~(\ref{eq:Bfunction}) and $b_i = (a_i + 1)/2$. The prefactor $M$ can be absorbed into the observable, as long as it is independent from $\vec{a}$. 
To certify equality, we will show
\begin{enumerate}
    \item that the cases (b), (d) and (e) vanish for all choices of $k,l,m,n$,
    as they contain odd numbers,
    \item that the results of all integrals in the family (a) coincide,
    as well as those in family (c), as the function $B$ is symmetric w.r.t.~its parameters,
    \item that the relative factor between the results in family (a) and
    those in family (c) is given by $3$. This comes from the fact that $B(\frac52,\frac12,\ldots,\frac12) = 3 B(\frac32,\frac32,\frac12,\ldots,\frac12)$.
\end{enumerate}
With the help of the reference \cite{Puchala2017},
we analytically evaluate the five families
case by case, treating the eigenvalue of $y$ as a free variable
and starting with case (a).

\textit{Case (a).}
Depending on the value of $k$, s.t.~$a_k=4$, we obtain as a result of
the integral either one of the polynomials
\begin{align}
    P_1 &= C\left(\frac{(d+2)^2(d+3)}{32} - \frac{(d+1)(d+3)}{4}y + \frac{d^2+8d+3}{4}y^2 - 2dy^3 + dy^4\right), \tag{S.92} \\\nonumber
    P_2 &= C\left(\frac{(d+2)^2(d+2)}{32} - \frac{(d+1)(d+2)}{4}y + \frac{d^2+9d-6}{4}y^2 - (3d-4)y^3 + \frac{3d-4}{2}y^4\right) \\
        &+ \frac{C}{d-1}\left(\frac{d+1}{4} - 2y + 3y^2 - 2y^3 + y^4 \right), \tag{S.93}
\end{align}
or linear combinations of them with prefactors adding to one.
Setting $P_1=P_2$, we obtain the two real solutions
for $y$ given by Eq.~(\ref{eq:ysoldodd}).

\textit{Case (b).}
Depending on $k$ and $l$, there are two types of integrals:
One vanishes directly, the other yields a multiple of $P_1-P_2$,
which vanishes for our choice of $y$.

\textit{Case (c).}
This case yields a couple of different results, all of them given by linear
combinations of $P_1$ and $P_2$ with prefactors adding to $1/3$.
Substituting the solution for $y$, we obtain in every case the same result,
given by $1/3$ of the result obtained in case (a).

\textit{Cases (d) and (e).}
These cases are analogous to case (b), yielding zero in each case for the
obtained solution of $y$.

All together, we have shown that for the observable $M_d$ in odd dimensions
with $y$ given by Eq.~(\ref{eq:ysoldodd}), the fourth moment of
random unitary measurements coincides with that of random orthogonal ones. 

Finally, we consider the second moments. First, it has been shown in Ref.~\cite{Tran2016} that the second moments do not depend on the eigenvalues, as long as the observable is traceless. Then, note that the result given in Theorem 2 of Ref.~\cite{Tran2016}
also holds for mixed states, the proof given there directly applies to the mixed 
state case. This theorem states that the second moments $\RR^{(2)}$  have the same
expression as the one we derived for the second moments $\SSS^{(2)}$
in Eq.~(\ref{evaluateds2}).  So the claim follows.
\end{proof}

A similar result can be obtained for even dimensions. However, the solution
for $y$ is in this case less aesthetic.

\noindent
{\bf Observation 10.}
{\it
In $d$-dimensional quantum systems where $d$ is even,
let the diagonal observable $M_d$ be given by
\begin{align}
    M_d = \operatorname{diag}(\underbrace{\alpha^\prime_+,\ldots,\alpha^\prime_+}_{(d-2)/2},\beta^\prime_y,\underbrace{\alpha^\prime_-,\ldots,\alpha^\prime_-}_{d/2}), \tag{S.94}
\end{align}
where
\begin{align}
\alpha^\prime_\pm &= \frac{\pm 2d - 4(y-1)}{\sqrt{d^3 + 4d^2 (y-1)y - 4 d (y-1)^2}}, \tag{S.95}\\
\beta^\prime_y &= \frac{ 2d (2 y-1) - 4(y-1)}{\sqrt{d^3 + 4d^2 (y-1)y - 4 d (y-1)^2}}, \tag{S.96}
\end{align}
and $y$ is obtained by the real solutions to $P^\prime_1 = P^\prime_2$ of
\begin{align}
    P^\prime_1 &= C^\prime\left(\frac{d(d+2)^2}{32} - \frac{(d+2)^2}{4}y + \frac{(d+1)(d+6)}{4}y^2 - 2dy^3 + (d-1)y^4\right), \tag{S.97} \\\nonumber
    P^\prime_2 &= C^\prime\left(\frac{(d^2+3d+3)d}{32} - \frac{d^2+3d+3}{4}y + \frac{d^2+8d+2}{4}y^2 - (3d-4)y^3 + \frac{3d-7}{2}y^4\right) \\ 
        &+ \frac{C^\prime}{d-1}\left(\frac{3d}{32} - \frac34y + \frac12y^2 - 2y^3 + 3\frac{d-1}{d}y^4 \right), \tag{S.98}
\end{align}
Then, measuring the observable $M_d$
also yields the coincidence between $\mathcal{R}^{(r)}_{AB}$ and $\mathcal{S}^{(r)}_{AB}$,
for $r=2,4$.
}

\begin{proof}
The proof follows exactly the same lines as those of
Observation 9.
\end{proof}

\begin{figure}
    \centering
    \includegraphics[width=0.5\columnwidth]{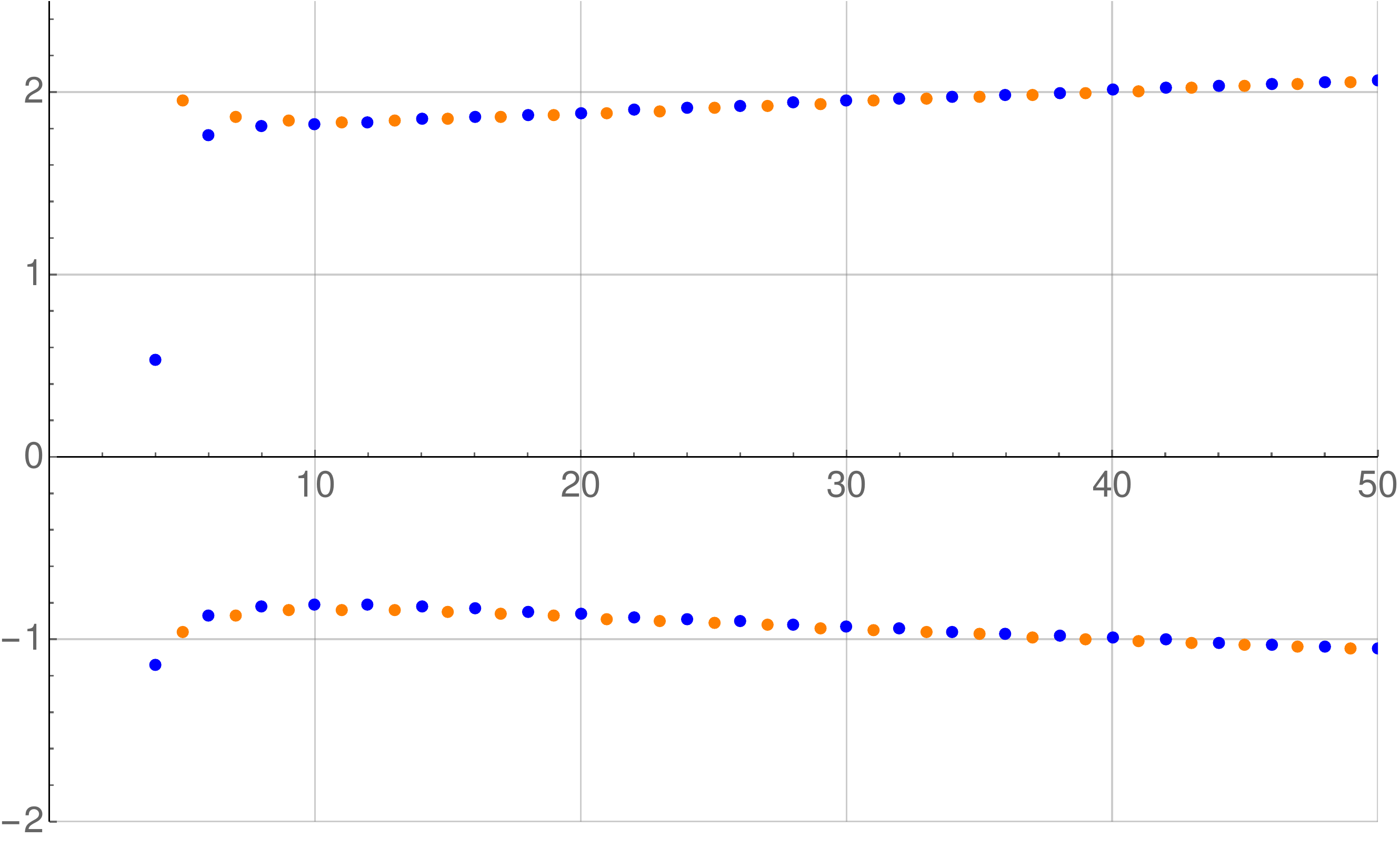}
    \caption{
    Values of $y$ such that the unitary integral yields the same value as the orthogonal one.}
    \label{fig:obs_sol}
\end{figure}

\begin{remark}
For reference, the solutions for $y$ for odd and even dimensions are plotted in Fig.~(\ref{fig:obs_sol}).
\end{remark}

\subsection*{B3. Maximization and minimization of the fourth moment $\mathcal{S}_{AB}^{(4)}$}
Let us discuss the maximization and minimization of the fourth moment
$\mathcal{S}_{AB}^{(4)}$ for separable states,
which leads to the area plotted in Fig.~\ref{3by3figure}.
As we described in the main text, the
moments $\mathcal{S}^{(r)}_{AB}$ are, by definition, invariant under 
orthogonal transformations of the submatrix $T_{\mathrm s}$,
where $T_{\mathrm s}=(t_{ij})$ with $1 \leq i,j \leq d^2-1$ in
Eq.~(\ref{eq-twoqudits}).
This orthogonal invariance allows us to consider the diagonalization of
the submatrix $T_{\mathrm s}$
\begin{align}
    T_{\mathrm s}^{\prime} = O_A T_{\mathrm s} O_B^{\top}=\mathrm{diag}(\tau_1,\tau_2, \ldots, \tau_{d^2-1}),
    \tag{S.99}
\end{align}
where $O_A, O_B \in SO(d^2-1)$ are non-physical orthogonal matrices
and $\tau_i\geq0$ are singular values of $T_{\mathrm s}$.
With this, we are able to reduce the number of parameters for
the moments $\mathcal{S}^{(r)}_{AB}$.
In fact, the evaluated second and fourth moments
(\ref{evaluateds2}, \ref{evaluateds4}) can be simply expressed as
\begin{align} 
    \mathcal{S}^{(2)}_{AB} 
&= V\sum_{i=1}^{d^2-1} \tau_{i}^2, \tag{S.100} \\
   \mathcal{S}^{(4)}_{AB} 
&=  W\left[ 2\sum_{i=1}^{d^2-1} \tau_{i}^4 + \left(\sum_{i=1}^{d^2-1} \tau_{i}^2 \right)^2 \right]
=W\left[ 2\sum_{i=1}^{d^2-1} \tau_{i}^4 + \frac{1}{V^2}
\left(\mathcal{S}^{(2)}_{AB}\right)^2 \right], \tag{S.101}
\end{align}
where $V=1/(d-1)^2$ and $W=1/3(d-1)^4$.
To maximize and minimize the fourth moment $\mathcal{S}^{(4)}_{AB}$,
we fix the second moment $\mathcal{S}^{(2)}_{AB}$
and employ the dV criterion as the constraint:
\begin{align}
    \|T_{\mathrm s}\|_{\text{tr}} = \|T_{\mathrm s}^{\prime}\|_{\text{tr}}
    =\sum_{i=1}^{d^2-1} \tau_i \leq d-1. \tag{S.102}
\end{align}
Then the task reads
\begin{align}
\max_{\tau_i}/\min_{\tau_i}
\ \ \ \ \ \ \ \ \ \ \ \ \ \ \ 
\mathcal{S}^{(4)}_{AB}&=W\left[ 2\sum_{i=1}^{d^2-1} \tau_{i}^4 +
\frac{1}{V^2}
\left(\mathcal{S}^{(2)}_{AB}\right)^2 \right], \tag{S.103}\\
\mathrm{s.t.}
\ \ \ \ \ \ \ \ \ \ \ \ \ \ \ \ \, \ \ \ \ \ \ \ 
\mathcal{S}^{(2)}_{AB} &= V\sum_{i=1}^{d^2-1} \tau_{i}^2, \tag{S.104}\\
\sum_{i=1}^{d^2-1} \tau_i &\leq d-1, \tag{S.105}\\
0 &\leq \tau_i \leq d-1, \tag{S.106}
\end{align}
where $\tau_i$ is maximal and equal to $d-1$
for a pure product state, 
due to the positivity of states.
Consequently, we can characterize the set of admissible values
$\left(\mathcal{S}^{(2)}_{AB},\, \mathcal{S}^{(4)}_{AB}\right)$
for separable states by maximizing and minimizing
the fourth moment.
If a state lies outside this set, then it must be entangled.

The lower bound for PPT states can be obtained by numerical optimization.
Also, the lower bound for general states can be
obtained by imposing the constraint
$\sum_{i=1}^{d^2-1} \tau_i \leq d^2-1$,
and the isotropic state
$\varrho_{\mathrm{iso}}$ in
Eq.~(\ref{isotrpoicstatedbyd})
satisfies the bound, which
proves it is optimal.

\begin{figure}[t]
    \centering
    \includegraphics[width=0.7\columnwidth]{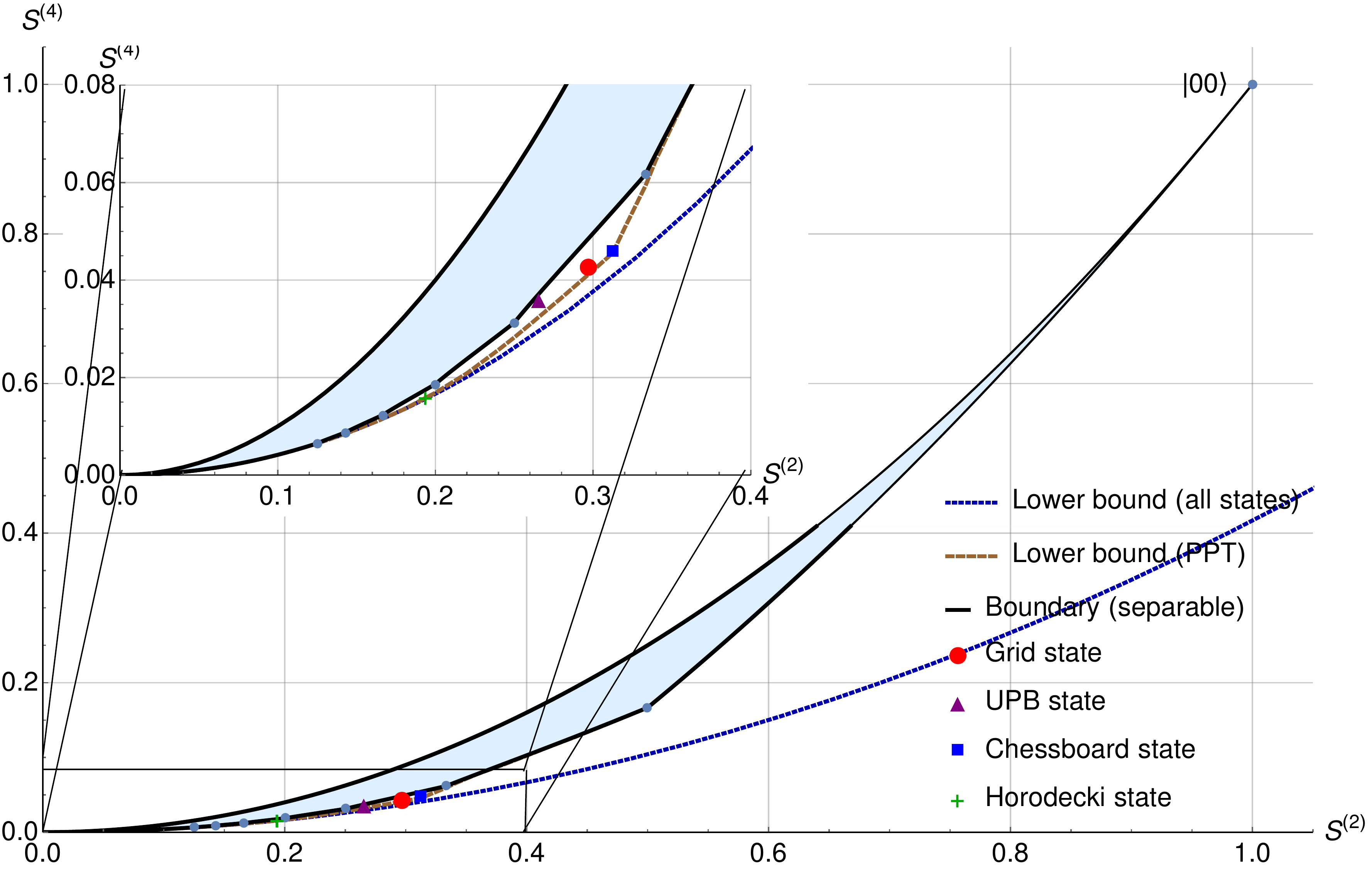}
    \caption{Entanglement criterion based on second and fourth 
    moments of randomized measurements for $3\otimes 3$ systems. }
    \label{3by3zoomoutfigure}
\end{figure}

\begin{figure}[t]
    \centering
    \includegraphics[width=0.7\columnwidth]{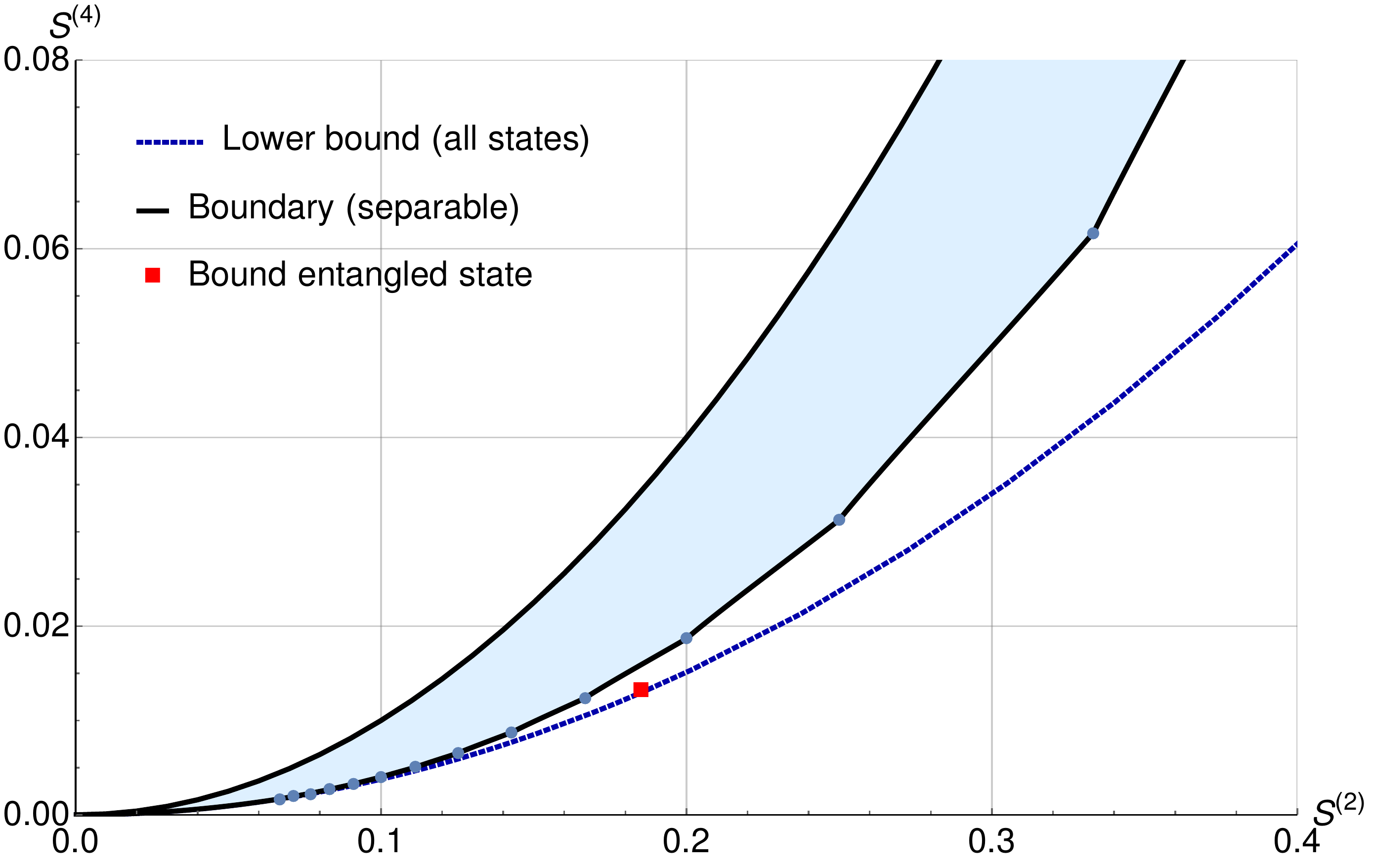}
    \caption{Entanglement criterion based on second and fourth 
    moments of randomized measurements for $4\otimes 4$ systems. The bound entangled state is outside, meaning that it can be detected with the methods
    developed in this paper.}
    \label{4by4zoomin}
\end{figure}

In the main text, we showed the zoom-in plot in
Fig.~\ref{3by3figure}.
Here, we also give the zoom-out plot in Fig.~\ref{3by3zoomoutfigure}.
Moreover, for reference, the results for $d=4$ are shown in
Fig.~\ref{4by4zoomin}.
Importantly, the $4\otimes4$ bound entangled Piani state
from the Refs.~\cite{Benatti2004} is
outside of the region
of separable states, meaning that
it can be detected by the method of moments with random correlations developed in this paper.

Let us discuss which states are good candidates for violating
our criterion. It is known that in $d\otimes d$-dimensional systems,
if the states have maximally mixed subsystems, then the dV criterion 
is equivalent to the CCNR criterion. If not, the dV criterion is weaker 
than the CCNR criterion (see Ref.~\cite{deVicente2008}). On the other hand, 
if an entangled state is very closed to a state with maximally mixed 
subsystems and largely violates the CCNR criterion, then we may detect 
the entangled state based on the dV criterion. For instance, the so-called 
cross-hatch $3\times3$ grid state, one of the bound entangled states
detected by our methods, does not have maximally mixed subsystems.
Nevertheless, its reduced states is close to maximally mixed, 
$\rho_A =\rho_B= \text{diag}(0.375, 0.25, 0.375)$, and moreover, it 
violates the CCNR criterion by a large amount.

\newpage
\section*{Appendix C: Bound entangled states}
In Appendix C, we give explicit forms of the
bound entangled states that can be detected by 
the methods with randomized measurements:
(1) quantum grid states
(2) chessboard states
(3) states forming unextendible product bases
(4) $3 \otimes 3$ Horodecki state
(5) the $4 \otimes 4$ bound entangled Piani state.
We remark that Yu-Oh bound entangled states \cite{Yu2017} and
steering bound entangled states \cite{TMoroder2014}
cannot be detected by our methods.
\subsection*{C1. Quantum grid states}
A quantum grid state is a toy model that can describe the
mixture of entangled states.
Its entanglement properties have been characterized using entanglement criteria
\cite{Lockhart2018}.
For $d \otimes d$-dimensional systems, consider a pure entangled state forming
\begin{align}
    \ket{i,j;k,l}=\frac{1}{\sqrt{2}}\left(\ket{ij}-\ket{kl} \right), \tag{S.107}
\end{align}
with $0\leq i,j,k,l <d$.
A quantum $d\times d$ grid state is defined as
the uniform mixture of pure states $\ket{i,j;k,l}$.
That is, for a given set $E=\{\ket{i,j;k,l}\}$, it can be defined as
\begin{align}
\varrho(E)=    \frac{1}{|E|}\sum_{\ket{e}\in E}\ket{e}\!\bra{e}. \tag{S.108}
\end{align}
Note that not all quantum grid states are separable,
and moreover, there are some grid states that can have bound entanglement.
In particular, a $3\times 3$ bound entangled grid state is
called the cross-hatch state with the set $E_{\mathrm{ch}}=\{\ket{1,1;2,3}, \ket{2,1;3,3}, \ket{1,2;3,1}, \ket{1,3;3,2}\}$, see also Fig.~2 (a) in \cite{Lockhart2018}.
It is known that the cross-hatch state is detected by the CCNR criterion.
\subsection*{C2. Chessboard states}
For $3 \otimes 3$-dimensional systems, consider a family of quantum states
\begin{align}
    \varrho_{\mathrm{cb}} = N\sum_{i=1}^{4}\ket{V_i}\!\bra{V_i}, \tag{S.109}
\end{align}
where $N=1/\sum_i \Braket{V_i|V_i}$ is a normalization factor and
\begin{align}
    \ket{V_1}&=(m,0,s;0,n,0;0,0,0), \tag{S.110}
    \\
    \ket{V_2}&=(0,a,0;b,0,c;0,0,0), \tag{S.111}
    \\
    \ket{V_3}&=(n,0,0;0,-m,0;t,0,0), \tag{S.112}
    \\
    \ket{V_4}&=(0,b,0;-a,0,0;0,d,0). \tag{S.113}
\end{align}
with free real parameters $a,b,c,d,m,n$ and $s=\frac{ac}{n}, t=\frac{ad}{m}$. 
The matrix form of this state can then be expressed as
\begin{align}
\varrho_{\mathrm{cb}}= 
\begin{pmatrix}
m^2+n^2&0&ms&0&0&0&nt&0&0 \\ 
0&a^2+b^2&0&0&0&ac&0&bd&0 \\
sm&0&s^2&0&sn&0&0&0&0 \\ 
0&0&0&a^2+b^2&0&bc&0&-ad&0 \\
0&0&ns&0&m^2+n^2&0&-mt&0&0 \\ 
0&ac&0&cb&0&c^2&0&0&0 \\ 
tn&0&0&0&-tm&0&t^2&0&0 \\
0&bd&0&-da&0&0&0&d^2&0 \\ 
0&0&0&0&0&0&0&0&0 \\ 
\end{pmatrix}. \tag{S.114}
\end{align}
The states $\varrho_{\mathrm{cb}}$ are called the chessboard states
because their $8\times8$ matrix form looks like a chessboard,
originally introduced by Dagmar Bru{\ss} and Asher Peres \cite{Bruss2000}.
The state $\varrho_{\mathrm{cb}}$ is invariant under the partial
transposition:
$\varrho_{\mathrm{cb}} =\varrho_{\mathrm{cb}}^{T_B} \geq 0$.
On the other hand, according to the range criterion \cite{PHorodecki1997},
$\varrho_{\mathrm{cb}}$ is entangled. Thus, the chessboard states are bound entangled. 
The extremal PPT entangled state shown in Fig.~\ref{3by3figure} in the main text and Fig.~\ref{3by3zoomoutfigure} is a state from this family with $m=n=b=-3/5$, $a=3/5$, $c=-d=6/5$.

\subsection*{C3. Unextendible product bases}
For $3 \otimes 3$-dimensional systems, consider five product states
\begin{align}
    \ket{\psi_0}&=\frac{1}{\sqrt{2}}\ket{0}(\ket{0}-\ket{1}),
    \ \ \ \ \ 
    \ket{\psi_1}=\frac{1}{\sqrt{2}}(\ket{0}-\ket{1})\ket{2},
    \ \ \ \ \ 
    \ket{\psi_2}=\frac{1}{\sqrt{2}}\ket{2}(\ket{1}-\ket{2}), \tag{S.115}
    \\
    \ket{\psi_3}&=\frac{1}{\sqrt{2}}(\ket{1}-\ket{2})\ket{0},
    \ \ \ \ \ 
    \ket{\psi_4}=\frac{1}{3}(\ket{0}+\ket{1}+\ket{2})(\ket{0}+\ket{1}+\ket{2}). \tag{S.116}
\end{align}
Notice that all of these five product states are orthogonal to
all pairs, and another product state cannot be orthogonal to all pairs.
These product states are said to form an unextendible product basis
(UPB) \cite{Bennett1999}. 
From these states, one can construct the mixed state
\begin{align}
    \varrho_{\mathrm{UPB}}=\frac{1}{4}\left(\eins - \sum_{i=0}^{4}\ket{\psi_i}\! \bra{\psi_i} \right). \tag{S.117}
\end{align}
Here, $\varrho_{\mathrm{UPB}}$ is the state on the space that is orthogonal to the space spanned by the UPB.
Then, $\varrho_{\mathrm{UPB}}$ has no product states in the range.
According to the range criterion \cite{PHorodecki1997},
$\varrho_{\mathrm{UPB}}$ should be entangled.
On the other hand, one can notice that $\varrho_{\mathrm{UPB}}$
is invariant under the partial
transposition:
$\varrho_{\mathrm{UPB}}^{T_B} = \varrho_{\mathrm{UPB}} \geq 0$.
Hence, $\varrho_{\mathrm{UPB}}$ is a bound entangled state.

\subsection*{C4. $3 \otimes3$ Horodecki state}
For $3 \otimes 3$-dimensional systems, consider the mixed state
\begin{align}
    \sigma(p)=\frac{2}{7}\ket{\psi^+}\!\bra{\psi^+}+\frac{p}{7}\sigma_+
    +\frac{5-p}{7}\sigma_-, \ \ \ \ \ \ \ 2\leq p \leq 5, \tag{S.118}
\end{align}
where
\begin{align}
    \ket{\psi^+}&=\frac{1}{\sqrt{3}}(\ket{00}+\ket{11}+\ket{22}), \tag{S.119}
    \\
    \sigma_+&=\frac{1}{3}(\ket{01}\!\bra{01}+\ket{12}\!\bra{12}+\ket{20}\!\bra{20}), \tag{S.120}
    \\
    \sigma_-&=\frac{1}{3}(\ket{10}\!\bra{10}+\ket{21}\!\bra{21}+\ket{02}\!\bra{02}). \tag{S.121}
\end{align}
It turns out that the state $\sigma(p)$ is PPT in the range $2\leq p \leq 4$.
To characterize this state further, one can employ a
non-decomposable positive map $\Lambda$
such that $(\eins \otimes \Lambda)\sigma \ngeq 0$.
An example is  
\begin{align}
    \Lambda
    \begin{pmatrix}
a_{11}&a_{12}&a_{13} \\ 
a_{21}&a_{22}&a_{23} \\ 
a_{31}&a_{32}&a_{33} \\ 
\end{pmatrix}
=
\begin{pmatrix}
a_{11}&-a_{12}&-a_{13} \\ 
-a_{21}&a_{22}&-a_{23} \\ 
-a_{31}&-a_{32}&a_{33} \\ 
\end{pmatrix}
+\begin{pmatrix}
a_{22}&0&0 \\ 
0&a_{33}&0 \\ 
0&0&a_{11} \\ 
\end{pmatrix}. \tag{S.122}
\end{align}
This non-decomposable map allows us to classify this state as follows \cite{PHorodecki1999}:
the state $\sigma(p)$ is not detected as entangled for $2\leq p \leq 3$, PPT (bound) entangled  for $3< p \leq 4$, and NPT entangled for $4< p \leq 5$.

\subsection*{C5. $4 \otimes4$ bound entangled Piani state}
For $4 \otimes 4$-dimensional systems, consider the
orthogonal projections
\begin{align}
    P_{ij}=\ket{{\Psi_{ij}}}\!\bra{{\Psi_{ij}}}, \tag{S.123}
\end{align}
where
$\ket{{\Psi_{ij}}}=(\eins \otimes \sigma_{ij})\ket{\Psi_+^4}$,
$\ket{\Psi_+^4}=\frac{1}{2}\sum_{k=0}^{3}\ket{kk}$,
and $\sigma_{ij} = \sigma_i \otimes \sigma_j$ with Pauli matrices.
With these projections, one can construct the state
\begin{align}
\varrho_{\mathrm{BE}}
&= \frac{1}{6}
\left(P_{02}+P_{11}+P_{23}+P_{31}+P_{32}+P_{33}
\right) \tag{S.124}\\
&=\frac{1}{6}\left(
\Phi^+_{AB}\Psi^-_{A^{\prime}B^{\prime}}
+\Psi^+_{AB}\Psi^+_{A^{\prime}B^{\prime}}
+\Psi^-_{AB}\Phi^-_{A^{\prime}B^{\prime}}
+\Phi^-_{AB}\Psi^+_{A^{\prime}B^{\prime}}
+\Phi^-_{AB}\Psi^-_{A^{\prime}B^{\prime}}
+\Phi^-_{AB}\Phi^-_{A^{\prime}B^{\prime}}
\right), \tag{S.125}
\end{align}
where $\Phi^+, \Phi^-, \Psi^+, \Psi^-$
are also projectors on the Bell states.
It has been shown that the state $\varrho_{\mathrm{BE}}$ is
the $4 \otimes 4$ bound entangled under
the bipartition of $AA^{\prime}|BB^{\prime}$
\cite{Benatti2004}.

\section*{Appendix D: Numerical methods}
In Appendix D, we provide numerical methods to check the results presented in the main text.
Indeed, our conjecture that Observation~2 holds for mixtures of product states with different bipartitions, as well as the PPT boundary in Fig.~\ref{3by3figure} are obtained by extensive numerical searches. These are performed using Python and the optimization functions of the package SciPy \cite{SciPy}.

In particular, when optimizing over mixed separable states to check Observation~2, we parametrize these states for a fixed rank $r$ by constructing $r$ unnormalized pure product states w.r.t.~different bipartitions. We mix these states and rescale the result to form a proper quantum state. The actual optimization is then performed using 
the BFGS algorithm \cite{broyden1970, fletcher1970, goldfarb1970, shanno1970}, implemented in SciPy. We ran the optimization for different values of $r$ up to $r=8$, and all possible choices of the bipartitions. In order to minimize the risk of getting stuck in local minima, we repeated each of the optimizations $50$ times with random initial parameters.

In order to obtain the PPT bound in Fig.~\ref{3by3figure}, we parametrize the random density matrices by rearranging the variables into a hermitian matrix, then form the normalized square of it to obtain a proper quantum state. We then minimize its fourth moment for different constant values of the second moment with the constraint that its partial transpose is positive. These constraints are implemented via penalty terms in the target function. 
We then sampled the range of the second moment and ran the optimization $10$ times for each value, to reduce the risk of being stuck in a local minimum. 

The source code for these optimizations is available upon reasonable request.

\twocolumngrid

\end{document}